\def\of{\begin{equation}}
\def\zf{\end{equation}}
\def\zff#1{{\label{#1}}\end{equation}}
\def\ofa{\begin{eqnarray}}
\def\zfa{\end{eqnarray}}
\def\zffa#1{{\label{#1}}\end{eqnarray}}
\def\nn{ \nonumber \\}
 
\def\jna#1{(\ref{#1})}

\def\frac#1#2{{#1 \over #2}}    
\def\tr{\hbox{Tr}}

\def\hbar{{\mathchar'26\mskip-9muh}}
\def\tilda#1{{\mathaccent "7E #1}}
\def\v#1{\mathaccent 20 #1}
 
\def\Ad{\hbox{ Ad }}

\def\d{d\kern-0.8 ex\vrule height 1.3 ex depth-1.24 ex width 0.7 ex 
\kern 0.15 ex}

\def\D{D\kern-1.7 ex\vrule height .87 ex depth-0.8 ex width 0.7 ex 
\kern 0.95 ex}

\documentstyle[eqsecnum,aps,twoside,amsfonts]{revtex}

\begin{document}

\draft

\sf

\tighten

\title{  New Approximations to the Fradkin representation
for Green's functions }

\bigskip
 
\author{ Predrag L. Stojkov}

\address{Physics Department, Brown University, Providence, RI 02912, USA
\\
Stojkov@het.Brown.edu}

\maketitle

\begin{abstract}

A new variant of the exact Fradkin representation of the Green's function
$G_c(x,y|gU)$, defined for arbitrary external potential $U$, is presented.
Although this new approach is very similar in spirit to that previously
derived
 by Fried and Gabellini, for certain calculations this specific variant,
with its prescribed approximations, is more readily utilizable. 
Application of the simplest of these forms is made to the $\lambda\Phi^4$
theory in four dimensions. 
 As an independent check of these approximate forms, an improved version of
the Schwinger-DeWitt asymptotic expansion of parametrix function is
derived.

\end{abstract}


\section{I\lowercase{ntroduction}}

 We here present a new approach for evaluating the (causal) Green's
function $G_c(x,y|gU)$ and the related loop-generating functional $L[gU]$,
for a particle in the presence of an arbitrary external field $U(x)$.  For
brevity's sake,
 the simplest case when both the propagating particle and the external
field are scalars is considered. The more interesting cases, when the
particle has spinor structure and/or when it moves in a background gauge
or gravitational field, will be presented in the next paper of this
series.

The causal Green's function for a scalar particle in the presence of a
scalar background $U(x)$ is a solution of
 the equation\footnote{Notation:  $a :\equiv b$ defines $a$ in terms of
$b$;  $\d p :\equiv \frac{d^Dp}{(2\pi)^D}$. ${\D \eta} :\equiv \Pi_{t=0}^s
\frac{d\eta(t)}{2\pi}$.  Metric has signature $(2-D)$: 
 $\eta_{**}=\eta^{**}= \hbox{diag}(+1, -1, \cdots, -1)$.  Also: $z:\equiv
x-y$; $\sigma(\xi)  :\equiv y + \xi z$;  $m^2_- :\equiv m^2 - i \epsilon$; 
$W_\mu(x):\equiv \partial_\mu U(x)$; ${\cal H}_{\mu\nu}(x)  :\equiv
\partial_\mu \partial_\nu U(x)$ ;  \\ We will use the {\bf hat} symbol for
operators that are of continuous nature (differential or integral) 
($\hat{K}$), the { \bf hacheck } symbol for matrices, ($\v{{\cal H}}$) 
and the {\bf bold face }) for objects that are both functional operators
and matrices. }

\of
[ \hat{P}^2_x -m^2_-   +  g U(x)  ]  G_c(x,y|gU)  = - \delta(x-y) ,
\zf

\noindent and can be presented in the form 

\of
G_c(x,y|gU)  =i\int_0^\infty ds  e^{-ism^2_-} g(s|x,y|gU)  ,
\zf

\noindent where $g(s|x,y|gU):\equiv e^{i s [\hat{P}^2_x+ g
U(x)]}\delta\left(x-y\right)$ is {\bf the Parametrix function} (with name
taken from ref. \cite{Hurt83}) of the problem.  Since its relation to both
Green's function and loop-generating functional are very close, we will
dedicate the major part of this paper to a derivation of different exact
and approximate forms of the parametrix function, and results from the
existing literature will be rephrased in terms of it.  There exist two
essentially different approaches for the construction of the parametrix:
the Fradkin forms and the Schwinger-DeWitt asymptotic expansion. The first
approach will be explained here, while the second is postponed until we
reach Section \ref{SchwingerDeWittAsymptoticExpansionSection}. 

The first approach is based on Fradkin's \cite{Fradkin66} exact
representation

\of
                 g(s|x,y|gU) = 
		e^{-i\int_0^{s} dt \delta_v^2} 
		e^{i g \int_0^{s} dt U\left(y + \int_0^{t} v\right)}
                \left. 
		\delta\left(x-y-\int_0^{s} v\right) 
		\right|_{v\rightarrow 0}
\zff{minkowskifradkinpropagator}

\noindent where $e^{-i\int_0^{s} dt \delta_v^2}$ denotes {\bf the linkage
operator} acting on the auxiliary variable $v(t)$ ($0\le t \le s$), and
$\delta_v^2:\equiv \left(\frac{\delta}{\delta v(t)}\right)^2$. This form,
on the one hand, leads directly to a path-integral formulation (Appendix
\ref{izvodenjefunkcinalneforme}), a transition based on equivalence
relations between the functional-linkage and functional integral forms
(Appendix \ref{linkageidentiteti}).  On the other hand, Fradkin`s
expression \jna{minkowskifradkinpropagator} allows a derivation of an
entire spectrum of different applications and approximations
(\cite{Fried1}, \cite{Fried2}.). The majority of these are eikonal
inspired, such as the {\bf Phase Averaged Approximation} ($<Ph>$) of
\cite{Fried93A}:

\of
   g^{<Ph>}(s|x,y|gU) = 
                \int \d p e^{ipz+isp^2}
                \int dq \d P e^{-i(q-y)(P+2p)}
                e^{ig\int_0^s dt U(q + t P)}
\zff{Phapproximacija}
 
\noindent  

It turns out that, based on the $<Ph>$-Approximation as an initial step,
it is possible to recast Fradkin's form \jna{minkowskifradkinpropagator}
in terms of Gaussian-weighted integrals over the set of auxiliary
4-vectors $\{Q_n,P_n|n=1,3,5,\cdots\}$, with the exact representation
(\cite{Fried93B},\cite{Fried94}):

\ofa
g(s|x,y|gU) &=& 
\int \d p e^{ipz+isp^2}
\int dq \d P e^{-iP(q-y)-i\frac14 s P^2}
{\prod_{n}}' i^2\int dQ_n\d P_n e^{-\frac{i}2[P_n^2+Q_n^2]}
\cr
& & \ \
e^{ig\int_0^s dt U\left(q+t(P-2p)- \frac{2\sqrt{s}}\pi {\sum_n}'
 \frac1n
\left[
P_n \cos\left(\frac{n\pi t}{s}\right)
+
Q_n \sin\left(\frac{n\pi t}{s}\right)
\right] 
\right)}
\zffa{minkowskifriedspropagator}

\noindent where the prime ($'$) denotes summation over odd natural
numbers, $\{Q_n,P_n\}$ are pairs of auxiliary 4-vectors.  The steps needed
to obtain this form are:  Taylor-expand the exponential $e^{ig \int_0^s dt
U\left(y+\int_0^s v\right)}$; use Fourier representations for each
$U$-factor and for $\delta\left(x-y-\int_0^{s} v\right)$;  and evaluate
the $v$-variable linkage operation. This generates exponentials with
binary products $k_i\cdot k_j$ of the Fourier momenta $k_j$ for different
$U$-factors, and these products are multiplied by functions of the
difference of corresponding auxiliary (proper) time variables $t_i-t_j$. 
One of these functions is $|t_i-t_j|$; and a suitable separation in terms
of its Fourier modes has to be used to allow the factorization of
expressions corresponding to different $U$ factors. Once that separation
is achieved, the reverse (inverse) Fourier transformation has to be
performed, and the resulting series of factorized terms must be resumed.
After some trivial rescalings, the resulting expression is
\jna{minkowskifriedspropagator}.

This form provides one with the possibility of approximating the full
(exact) expression by taking into account only several pairs of variables
$\{Q_n,P_n\}$.  The $<Ph>$ approximation is obtained by neglecting all
such $\{Q_n, P_n\}$ variables in the argument $U$ in
\jna{minkowskifriedspropagator}; also, the term $-\frac{i}4 s P^2$ should
be dropped \cite{Fried93B}.
 Taking into account only the first pair of auxiliary variables
$\{Q_1,P_1\}$ in expression \jna{minkowskifriedspropagator} gives an
improved ("$<Ph|1>$") approximation, etc ... It is possible to quantify
the quality of such approximations, in terms of the quality of
corresponding approximations to the $|t_i-t_j|$ function; and, according
to \cite{Fried93B}, the $<Ph|1>$ may be expected to carry $81\%$ of
contribution to the total expression, the $<Ph|1|3>$ $90\%$, the
$<Ph|1|3|5>$ $93\%$ etc ...

\section{N\lowercase{ew representation for parametrix}}

In the course of searching for an alternate derivation of
\jna{minkowskifriedspropagator} from Fradkin's form
\jna{minkowskifradkinpropagator} (without resorting to a perturbative-like
expansion and subsequent resummation), we have found a new, exact form,
based on the mode expansion:

\of
v(t)^\mu= V^\mu + \sum_{n=1}^\infty [Q_n^\mu \cos(n\omega t ) +P_n^\mu 
\sin(n\omega t)] 
\zff{obivniragdkgm}

\noindent where $\omega=\frac{2\pi}{s}$. The main criteria in developing
such an expansion is (the ad hoc) requirement that the delta-function
restriction $\int_0^s dt v(t) = z$ constrains only the amplitude of the
zero-frequency mode $V$ (forcing it to be $=\frac{z}{s}$), thus leaving
amplitudes of the oscillating modes unconstrained. The properly normalized
basis on the interval $[0,s]$ is $\left\{ \phi_n \right\}$ $=$ $\left\{
c_0(t) \equiv \sqrt{\frac1{s}} ; \right. $ $ \left. c_{n}(t) \equiv
\sqrt{\frac2{s}} \cos(n\omega t) ;\right. $ $ \left.  s_{n}(t) \equiv
\sqrt{\frac2{s}} \sin(n\omega t); \ (n\ge 1)
 \right\}$, where the scalar product is just the integral over that interval. 

With such a basis the resolution of the identity on the same interval is
provided:

\ofa
\delta(t-t') &\equiv& \sum_\lambda\phi_\lambda(t)^*\phi_\lambda(t') 
=  
\frac1{s} + \frac2{s} \sum_{n=1}^\infty \cos(n\omega (t-t')) ;
\zfa

\noindent and, requiring that $ {\delta v(t)^\mu \over \delta v(t')^\nu }
= \delta(t-t')  \delta^\mu_\nu $, the corresponding mode-decomposition of
the functional derivative\footnote{ For the general decomposition
$v(t)^\mu=\sum_\lambda V_\lambda^\mu \phi(t)_\lambda$ where
$\{\phi(t)_\lambda\}$ is an ortho-normal basis; the corresponding
functional derivative is given by $ {\delta \over \delta v(t)^\mu }
=\sum_\lambda \phi_\lambda(t)^* {\partial \over \partial V^\mu_\lambda}$.}
: 

\of
{\delta \over \delta v(t)^\mu } = \frac1{s} {\partial \over\partial 
V^\mu} +  \frac2{s} \sum_{n=1}^\infty  \left\{ \cos(n\omega t){\partial 
\over \partial Q^\mu_n}  + \sin(n\omega t)  {\partial
\over \partial P^\mu_n}\right\}
\zff{kakonaciizvod}

Using \jna{kakonaciizvod},  one can calculate some necessary expressions:

\ofa
\int_0^s dt {\delta^2 \over \delta v(t)^2 } 
&=&
\frac1{s} {\partial^2
\over \partial V^2} + \frac2{s} \sum_{n=1}^\infty 
\left\{ 
{\partial^2 \over \partial Q_n^2} +  
{\partial^2 \over \partial P_n^2} 
\right\} \ ,
\nn
\nn
\int_0^{t} dt' v(t') 
&=&
 V t + \frac{s}{2\pi} \sum_{n=1}^\infty\frac1{n} 
\left\{ Q_n \sin(n\omega t) + P_n [1- \cos(n\omega t)] \right\} 
\ ,
\nn
\nn
\int_0^s dt v(t) 
&=& 
V s \ ;
\zfa

\noindent and evaluate the $V$-linkage operation (using eq.
\jna{trecazgodna}), to
obtain the  new form for the parametrix
\jna{minkowskifradkinpropagator}:

\ofa
g(s|x,y|gU)  
&=& 
\frac{i}{(4\pi i s)^{\frac{D}2}}
e^{ - \frac{iz^2}{4s} }
 \prod_{n=1}^\infty                                   
e^{ - \frac{2i}{s}
\left\{
{\partial^2 \over \partial Q_n^2} +  {\partial^2 \over \partial P_n^2 }
\right\}
}
\nn
& &
\left.
e^{
i g \int_0^{s} dt
U\left( y + z \frac{t}{s} + \frac{s}{2\pi} \sum_{n=1}^\infty \frac1{n}
        \left\{ Q_n \sin(n\omega t) + P_n [1- \cos(n\omega t)] \right\}
\right)
}
\right|_{\{ Q_n \},\{ P_n \}\rightarrow 0 }
\zffa{tacnamoja}

This form is an exact representation of the Fradkin parametrix. Note that
the characteristic "frequencies" for modes in auxiliary space
$[0,s]$ employed here are even multiples of $\frac\pi{s}$, as
opposite to the FG form \jna{minkowskifriedspropagator}, where the
characteristic frequencies are odd multiples of that value. Also,
here the variables $V, \{P_m, Q_m\}$'s  are just Fourier mode  
of the auxiliary variable $v(t)$, while in form
\jna{minkowskifriedspropagator} no such simple connection between
$v(t)$ and their $\{P_m, Q_m\}$'s is known. 

In the following Section we will develop several approximate forms to
 \jna{tacnamoja}, as well as one suitable reformulation of that
exact result. That reformulation, in it's turn, then will give rise
to the some other approximate forms.

\section{$<0>$ A\lowercase{pproximation}}

We are now able to develop the hierarchy of approximate forms (in the same
spirit as those of the $<Ph|1|3|\cdots>$).  The basic one is obtained by
neglecting all oscillatory modes ($Q_n$`s and $P_n$`s) in the exponential
argument of $U$ in eq. \jna{tacnamoja}:

\of 
g^{<0>}(s|x,y|gU) = \frac{i}{(4\pi i s)^{\frac{D}2}} 
e^{ - \frac{iz^2}{4s} + i g \int_0^{s} dt U\left(
\sigma\left(\frac{t}{s}\right)\right) } 
\zff{NULAforma}

\noindent where $\sigma(\alpha) = y+ \alpha z$.  We will call this {\bf
the $<0>$-Approximation}. It has a simple meaning in coordinate space:
propagation is replaced by motion along the straight line between points
$x$ and $y$. In other words, the $<0>$-approximation ignores the
"structure" of the potential $U(x)$, paying no attention to it's "valleys"
and "hills". In terms of a propagating particle, this is the extreme
high-energy (UV) approximation; in terms of a background potential, it is
requirement that all its spatial derivatives are "small"  (i.e.
$|\partial_{\mu_1}\cdots\partial_{\mu_n} U| << \frac{|U|}{L^n}$, where the
characteristic length $L$ is $\sim s <v^2>^{1/2}$). 

It is worth noticing that the corresponding Green's function has the form 
of a free particle's Green's function with a 
path-dependent mass:

\of
\left.
G^{<0>}_c(x,y|gU)  = - \int_0^\infty ds \frac1{(4\pi i s)^{\frac{D}2}} 
e^{ -i s {\cal M}(x,y)^2_- - \frac{iz^2}{4s}} =  G_c(x,y|0) \right|_{m^2
\rightarrow
{\cal M}(x,y)^2 } 
\zff{<0>explicit}

\noindent where $ {\cal M}(x,y)^2 \equiv m^2 - g \int_0^1 d\alpha
U(\sigma(\alpha))$. From this form are evident the properties of symmetry,
under $x\leftrightarrow y$;  and of the correct free-particle limit,
$U\rightarrow 0$.
 
The $<0>$-approximation is not same as the $<Ph>$.  This can be seen by
comparing a transformed
 (as closely as possible to the $<Ph>$-form; see Appendix
\ref{appendixzanultu})  expression for $<0>$:

\of
g^{<0>}(x,y|gU) =
\int \d p 
e^{ipz + isp^2} 
\int dq \d P  e^{-iq(P+2p)}   e^{ ig \int_0^s dt  U(y+\bf{ \frac{t}{s} q}+
t P)} ,
\zff{kolikotijegodina}

\noindent with:

\of
g^{<Ph>}(x,y|gU) =
\int \d p 
e^{ipz + isp^2} \int dq \d P e^{-iq(P+2p)} e^{ ig  \int_0^s dt
U(y + {\bf 
q} +  t P)}  \zff{malenadevojcice}

\noindent The two forms are very similar; the only difference between them
is in the argument of the potential $U$: instead of $[y-q- t P]$ (in
$<Ph>$)  we have $[y-\frac{t}{s} q -t P]$ (in $<0>$). 

While the $<0>$-approximation has the free-particle-like coordinate form
\jna{<0>explicit} for the Green's function, no such simplification can be
made for the $<Ph>$-approximation. But, in momentum space picture, the
$<Ph>$-approximation has a nice interpretation \cite{Fried1}, while the
$<0>$ remains nontransparent.  Thus one may consider the
$<0>$-approximation better suited for coordinate-space applications, while
for diagrammatic applications, the $<Ph>$ seems better adapted.

One more property (besides $x\leftrightarrow y$ symmetry and the correct
free-particle limit) that is shared by both the $<0>$- and
$<Ph>$-approximation is their exactness for the plane-wave (laser-like) 
background $U(x)=\phi(k.x)$ (with $k^2=0$). The final form of the exact
Green's function of this field is the given by the same expression,
\jna{<0>explicit}, which is only the $<0>$-approximation for other
(non-laser)  backgrounds.  Also, it is worth noticing that in the family
of approximations based on forms \jna{kolikotijegodina} and
\jna{malenadevojcice}, where the argument of the potential $U$ is replaced
by $[y - q F_1(\alpha)- s P F_2(\alpha)]$, the laser-background case again
simplifies to the explicit $<0>$-form \jna{<0>explicit}.

\section{T\lowercase{he loop generating functional} $L[\lowercase{g}U]$ }

Another important object (besides the Green's functions)  for functional
field-theoretic calculations is the loop generating functional:

\of
L[gU] :\equiv -\frac12 \tr
\det\left[\frac{(\hat{P}_x^2-m_-^2+gU_x)}{(\hat{P}_x^2-m_-^2)}\right] 
= \frac12
\tr \det(G_c[gU]G_c[0]^{-1})
\zf

It can be expressed in the integral form $L[gU] = \frac12 \int_0^1
d\lambda \tr \left(G_c[\lambda gU]gU\right) = \frac12 \int_0^1 d\lambda
\int dx G_c(x,x|\lambda gU) gU(x)$.

In the $<0>$-Approximation \jna{<0>explicit} the Green's function has a
diagonal component of amount

\of
G_c^{<0>}(x,x|gU) = - \int_0^\infty ds \frac1{ (4\pi i s )^{\frac{D}2}} 
e^{-is(m^2_- -gU)} = \frac{-i}{(4\pi)^{\frac{D}2}}
(m^2_--gU)^{\frac{D}2-1} \Gamma\left(1-\frac{D}2\right)
\zf

\noindent and  one finds 
 
\ofa
L^{<0>}[gU] &=& 
\frac12 \int_0^1 d\lambda \int  dx
\frac{i}{(4\pi)^{\frac{D}2}}
(m^2-\lambda gU)^{\frac{D}2-1} gU \Gamma\left(1-\frac{D}2\right)
\cr
&=& \frac{i\Gamma\left(-\frac{D}2\right)}{2(4\pi)^{\frac{D}2}}
\int  dx \left[ (m^2-gU)^{\frac{D}2} - m^D \right]
\zffa{izrazasdlsfsfldetwerminantu}

This is a divergent expression for even $D$, with
$\Gamma\left(-\frac{D}2\right)$ expressing the UV-divergences. In what
follows, we will assume that $\Gamma\left(-\frac{D}2\right)$ is
regularized in some implicit way\footnote{One such regularization follows
from the definition of the $\Gamma$-function, by replacing it's UV-limit
($\rho=0$) with the finite value $M^2\Lambda^{-2}$:

\ofa
\Gamma\left(-\frac{D}2\right) 
 \sim
  \int_{M^2\Lambda^{-2}}^\infty
d \rho \rho^{-\frac{D}2-1} e^{-\rho}
 \sim  
  \int_{M^2\Lambda^{-2}}
d \rho \rho^{-\frac{D}2-1}
\sim
\frac2{D}\left(\frac\Lambda{M}\right)^D
\zffa{fffutyregfjgobmvnfhdgshsdjfkj}

\noindent where $M$ is some fixed mass scale. Therefore,
$\Gamma\left(-\frac{D}2\right) \sim \Lambda^D$
 }, allowing explicit evaluation of $<0>$-approximate forms of some simple
field theoretical models.

Note that for $D=2$ and $D=4$, expression
\jna{izrazasdlsfsfldetwerminantu} simplifies further to

\ofa
L^{<0>}[gU]_{D=2} 
&=&
-i \frac{g\Gamma(-1)}{8\pi} \int dx U(x)
\cr
L^{<0>}[gU]_{D=4}
&=&
i \frac{\Gamma(-2)}{32\pi^2} \int dx [-2m^2 gU(x) + g^2 U(x)^2]
\label{adadadadadkjkjkjkj}
\zfa

\section{A\lowercase{pplication to the} $\lambda\varphi^4$-\lowercase{model}}

\label{ovopoljepripadasvimavamaali}

To illustrate the $<0>$-approximate form, consider the $\lambda\varphi^4$
model in Minkowski $1+d$-dimensional space. Start from an action

\of
S = \int dx \left[
\frac12 (\partial\varphi)^2 - \frac12 m^2 \varphi^2 + \frac12 \mu^2 
\psi^2 + g 
\varphi^2 \psi \right]
\zf

\noindent which is (after evaluating the EOM of the field $\psi$) 
equivalent to the standard action with a potential term $V(\varphi)=-
\frac12 m^2 \varphi^2 - \frac{\lambda}4 \varphi^4$, where
$\lambda=2g^2/\mu^2$. This particular formulation is chosen in preparation
for subsequent steps where it is important to have the action with all
terms at most quadratic with respect to the any specific field (for
example, the interaction term $g \varphi^2 \psi$ is quadratic with respect
to the $\varphi$ field and linear with respect to the $\psi$ field). The
field $\psi$ may be thought of as the classical form of a composite
$\varphi^2$. 

The generating functional $Z[J,j]:\equiv \int [D\varphi D\psi]
exp\left\{-iS[\varphi,\psi] - i \int dx [\varphi J + \psi j]\right\} $ can
be transformed, following steps outlined in \cite{Fried2}, into the
form\footnote{An alternate form is $ Z[J,j] = \left.
e^{-\frac{i}{2\mu^2}\int \partial_\psi^2} e^{L[g\psi]-\frac{i}2\int J
\cdot \hat{G}_c[g\psi]\cdot J} \right|_{\psi=-\frac1{\mu^2}j} $ }

\of   
Z[J,j] = e^{L[2gi\delta_j]} e^{-\frac{i}2\int J \cdot 
\hat{G}_c[2gi\delta_j] \cdot J}
e^{\frac{i}{2\mu^2}\int j^2}
\zf 

\noindent where $\hat{G}_c[gU]:\equiv (\partial^2 + m^2_- - gU)^{-1}$ is
the causal Green's function for the field $\varphi$ in an external
potential $U$, and $L[gU]:\equiv \frac12
\ln\det\left(\hat{G}_c[gU]\right)$ is the corresponding vacuum-loop
generating functional.

In this section, the the $2n$-point quantum correlation functions

\ofa
\Gamma_{2n}(x_1,\cdots x_{2n}) 
& :\equiv &
 <(\varphi(x_1)\cdots \varphi(x_{2n}))_+> 
\equiv 
\left. \frac1{Z[0,0]}i^{2n} \frac{\delta^{2n}Z}{\delta J(x_1)\cdots \delta
J(x_{2n})}\right|_{J,j=0}
\cr
&=& 
i^n  
\left.
\left[ 
G_c(x_1,x_2|2gi\delta_j)\cdots
G_c(x_{2n-1},x_{2n}|2gi\delta_j) 
+ \hbox{permutations} 
\right] {\cal Z}[j]
\right|_{j=0}
\zfa 

\noindent will be evaluated in the 
$<0>$-approximation.   
Here {\small 
${\cal Z}[j]:\equiv  Z[0,j]/Z[0,0] = 
Z[0,0]^{-1}   exp\left\{L[2gi\delta_j]\right\}
exp\left\{\frac{i}{2\mu^2}\int j^2\right\} 
$}. 
Only the first of these terms  will be observed,
having all 
others related to this one by a permutations of coordinate
arguments $\{x_1,\cdots,x_{2n}\}$.

For $D=4$ use \jna{adadadadadkjkjkjkj} and \jna{drugazgodnao} to obtain

\of
{\cal Z}^{<0>}[j] =  
e^{
\frac{i}{2\mu^2 (1-a)}
\int dx \left( j^2 + \frac{2 a g m^2}{\lambda} j \right)
}
\zf

\noindent where $a:\equiv \frac{\Gamma(-2)\lambda}{8\pi^2}$. From
here, use \jna{<0>explicit} 

\ofa
\Gamma_{2n}(x_1,\cdots x_{2n}) 
& = &
(-i)^n
\prod_{j=1}^n \int_0^\infty ds_j  \frac1{(4\pi i s_j)^{\frac{D}2}}
e^{ - i \sum_j \frac1{4s_j}(x_{2j-1}-x_{2j})^2 - im^2_-  \sum_j s_j }
\cr 
& &
{\cal Z}^{<0>}\left[-  2g \sum_j s_j \int_0^1 d\alpha
\delta(y-x_{2j}
- \alpha (x_{2j-1} -x_{2j}))\right] 
\zfa

\noindent The ${\cal Z}^{<0>}$ factor can be represented as a product of two
factors 
$K_1 K_2$ where

\ofa
K1 
&:\equiv &
e^{
\left(
\frac{i\lambda}{1-a}
\right)
\sum_{l=1}^n \sum_{r=1}^n s_l s_r \int_0^1 d\alpha  
\int_0^1 d\beta \delta(x_{2l}-x_{2r}
+ \alpha (x_{2l-1} -x_{2l}) -\beta
(x_{2r-1} -x_{2r}))
}
\nn
K2  &:\equiv &
e^{ 
- i \frac{a m^2}{1-a} 
\sum_{l=1}^n s_l
}
\zfa

The exponential in $K_1$ has two sources of singularities:  First, in
"diagonal"  terms $ l=r$, the strongly UV-singular integrals $ \int_0^1
d\alpha \int_0^1 d\beta \delta_D((\alpha-\beta)(x_{2r-1}-x_{2r}))  $ is
present. It can be transformed to the form $ \int \d k
\left(\left(\xi_{2r-1}\right)^{-1} \sin \xi_{2r-1} \right)^2 $, where
$\xi_{2r-1}:\equiv \frac12 k\cdot(x_{2r-1}-x_{2r})$, that it is
well-behaved only in one direction (parallel to the four-vector
$(x_{2r-1}-x_{2r})^\mu$). This gives an overall UV-behaviour $ \sim
\Lambda^3 $, where $\Lambda$ is UV-cut-off parameter, the same as in
\jna{fffutyregfjgobmvnfhdgshsdjfkj}. 

A second source of singularities in $K_1$ are nondiagonal terms in case of
intersecting linear trajectories;  such trajectories occur whenever there
exist $\alpha$ and $\beta$ such that $ x_{2l} + \alpha (x_{2l-1}-x_{2l}) =
x_{2r} + \beta (x_{2r-1}-x_{2r})$. Their contribution $ \int_0^1 d\alpha
\int_0^1 d\beta \delta_D(x_{2l} + \alpha (x_{2l-1}-x_{2l}) - x_{2r} -
\beta (x_{2r-1}-x_{2r})) $
 can be transformed into the form 
$ \int \d k 
\left(\xi_{2r-1}\xi_{2l-1}\right)^{-1} 
\sin \xi_{2r-1} \sin \xi_{2l-1}
exp\left\{ \frac{i}2 \left(x_{2l}+x_{2l-1}-x_{2r}-x_{2r-1}\right)\cdot k
\right\}
$.
 From that expression one can infer that two $k$-space directions are
convergent, and in one more the integration is made convergent by a
fast-oscillating exponential factor. The remaining unregularized direction
gives an overall $\Lambda^1$ UV-behaviour for such intersecting
trajectories.

Then, since $a \sim \Gamma(-2) \sim \Lambda^4$, terms in the exponential
of $K_1$ will display one of three UV-behaviours: 
$\Lambda^{3-4}=\Lambda^{-1}$ (diagonal singular terms),
$\Lambda^{1-4}=\Lambda^{-3}$ (nondiagonal singular terms) or
$\Lambda^{-4}$ (nondiagonal nonsingular terms, i.e. nonintersecting
trajectories). In this way, in the $\Lambda\rightarrow \infty$ limit, one
obtains $K_1=1$.

Notice that in the $\Lambda\rightarrow \infty$ limit, the value of
bare-coupling constant $\lambda$ does not play any role, since it cancels: 
$\tilda{\lambda}=\frac\lambda{1-a} \rightarrow
 - \frac{ 8\pi^2}{\Gamma(-2)} \sim
\Lambda^{-4} \rightarrow 0$.  Only way to make $\tilda{\lambda}$ nonzero 
(and finite) in the $\Lambda \rightarrow \infty$ limit is to accept peculiar
behaviour of bare coupling $\lambda =
\frac{\tilda{\lambda}}{1-\frac{\tilda{\lambda}\Gamma(-2)}{8\pi^2}} \sim
\Lambda^{-4} \rightarrow 0$, which
 is comletely opposite to the usual behaviour of bare couplings in the
Abelian models.

Concerning the factor $K_2$, there are no other (than $a$) singularly
behaving quantities in the exponential, and one can take the
$\Lambda\rightarrow \infty$ limit right away. The result is $K_2 = e^{+
im^2 \sum_{l=1}^n s_l}$. Or, since is not necessary to take that limit at
this point, one can join it to the bare-mass terms in the expression for
the $\Gamma_{2n}$, having, in effect, replaced the
 bare-mass $m^2$ with $M^2 :\equiv m^2 + \frac{a m^2}{1-a} = \frac1{1-a}
m^2$ in all subsequent expressions.

Then the $2n$-point correlator factorizes:

\of
\Gamma_{2n}(x_1,\cdots x_{2n}) \rightarrow 
\left. \prod_{j=1}^n G_c(x_{2j-1},x_{2j}|0)\right|_{m^2\rightarrow
M^2}
\zf

\noindent into product of a free-particle Green functions.

Alternatively, one can evaluate the generating functional in naive
$\Gamma(-2) \rightarrow \infty$ limit to get

\of
Z^{<0>}[J,0] \rightarrow  
Z^{<0>}[0,0] 
e^{\frac{i}2 \int J G^{<0>}_c\left[- a M^2\right]J} =
Z^{<0>}[0,0] 
e^{\frac{i}2 \int J 
G_c[0]|_{m^2\rightarrow M^2} J} =
Z[J,0]|_{\lambda, m^2\rightarrow M^2} 
\zff{ovojevidenekoflepo}

\noindent i.e. such a limit produces a free theory of fields of mass $M$.
Notice that it is not necessary to take the $<0>$-approximate form for
$G_c[gU]$ in \jna{ovojevidenekoflepo}: simplification in the $K_1$ factor
in ${\cal Z}^{<0>}[j]$ suffices to validate the above expression with the
exact $G_c[gU]$'s.

The disappearance of the effective interaction in the UV-limit of
$\varphi^4$ theory in four dimensions is the well-known phenomena of {\bf
triviality}, established or indicated in several different frameworks
(Euclidean constructive field theory:  \cite{Aizenman81}-\cite{Grosse88}; 
Lattice calculations and "inspired"  perturbative-renorm-group
extrapolations:  \cite{Agodi 94A}-\cite{Consoli 93} ; Functional approach: 
\cite{Fried80}).  Here, we have presented a simple analytic derivation of
triviality.  However, this method is not uniformly suitable for all number
of dimensions: for example, in $D=2$ the quadratic linkage operation is
absent, leading to the absence of UV-singular quantities in denominators
(that were of such importance in the $D=4$ case), and thus to the
non-triviality of the theory.  In $D\ne 2, 4$, the linkage functional
\jna{fffutyregfjgobmvnfhdgshsdjfkj} is either (in case of even $D$)  a
polynomial of order higher than two, or a nonpolynomial (irrational)
function (in the case of odd $D$) of $gU$; both circumstances generate
 linkage operators $e^{L[2ig\delta_j]}$ that are neither shift-operators
nor Gaussian linkages, preventing further evaluation in the absence of
additional approximate steps.

The more detailed discussion, concerning more uniform regularization of
sources UV-singularities, how this influence the derived trivial
behaviour, and corrections due to higher approximations (using $<L>$,$<Q>$
and approximations obtained using the Schwinger-DeWitt's approach; all
these are introduced in next few sections) will be presented elsewhere
\cite{Stojkov 97}.

\section{B\lowercase{eyond the} $<0>$-A\lowercase{pproximation}}

How may one proceed beyond the $<0>$-approximation? The goal here is to
obtain a sequentially better approximation to expression \jna{tacnamoja}. 
Imitating the $<Ph|1|3|\cdots>$-hierarchy does not help much, since the
potential $U(x)$ is in general a nonlinear function of $x$: for example,
retaining only $n=1$ level variables $Q_1$ and $P_1$ (the
$<0|1>$-Approximation) gives: 

{\small

\of
g^{<1>}(s|x,y|gU) = g(s|x,y|0)
e^{ -  \frac{2 i}{s}
\left\{ {\partial \over \partial Q_1^2} +  {\partial \over \partial P_1^2 }
\right\} }
\left.   
e^{ i g s
\int_0^1 d\alpha
 U\left(\sigma(\alpha)+ \frac{s}{2\pi}\left\{ Q_1 \sin(2n\pi  \alpha) +
P_1 [1- \cos(2n\pi  \alpha)] \right\}
 \right) }\right|_{Q_1 ,P_1\rightarrow 0 }
\zf

}

This expression can be evaluated analytically only in some special cases:
e.g. if the potential $U(x)$ is a quadratic polynomial of its arguments,
or if it represents a laser field (when it reduces to
$<0>$-approximation). For all other cases one has to resort to some kind
of further approximation, such as a
 Gaussian approximation. The same situation holds for $<0|1|2>$ and all
higher approximations. 

This indicates that it is perhaps not beneficial to insist on an
application of
 the hierarchy $<0|1|2|\cdots>$; but, instead, to proceed by an alternate
route. Let us, in $<0>$, "remove" the variable $V$ from $v(t)$, while
keeping all other components of $v(t)$ grouped into a new variable
$u(t):\equiv v(t)-V$. The variational derivative $\delta_{v}$ then splits
into $\delta_{v(t)^\mu} = \frac1s \partial_{V^\mu} + \delta'_{u(t)^\mu} $.
Here the $\delta'_{u(t)^\mu}$ is a {\bf constrained variational
derivative} that takes into account the constraint $\int_0^s dt u(t)^\mu
\equiv 0$; its action on $u(t')^\nu$ is $\frac{\delta u(t')^\nu}{\delta
u(t)^\mu} = \eta_\mu{}^\nu \Pi(t,t')$, where
$\Pi(t,t')=\left[\delta(t-t')-\frac1s\right]$ is a projector on the
$u$-subspace. In this way, one can write $\delta'_u = \Pi \delta_u$ in
terms of the unconstrained variational derivative $\delta_u$.

Then, the exact expression for the parametrix can be reformulated
(Appendix \ref{friedovdokazzamojidentitet}) as

\of
g(s|x,y|gU) = \frac{i}{(4\pi is)^{\frac{D}2}}
e^{- i\frac{z^2}{4 s}}
e^{ - i \int_0^s {\delta'}_u^2}
\left.
e^{i g \int_0^{s} dt  U\left(\sigma(\frac{t}{s})  + \int_0^t u\right) }
\right|_{u \rightarrow 0 }
\zff{tacniparametrix}.

While the $V$-integration (linkage operation) was easily performed in an
explicit manner (using identity \jna{trecazgodna}), the $u$-integration
cannot be done in an exact way.  Instead, one can assume that the $u$
variable is a "small"  quantity\footnote{A similar approach is called {\bf
the string inspired formalism} by Bern-Kosower \cite{Bern 92},\cite{Bern
93}:  the zero-mode variable is separated and the expansion over the
oscillating modes is performed, allowing the efficient resummation of many
parts of perturbative Feynman graphs}, thus generating a Taylor expansion
of the $U\left(\sigma(t/s) + \int_0^t u\right) $ in the $u$ variable, and
keeping only some of the lowest order terms. In what follows, derivation
of the two such lowest order approximations is defined by retaining only
up to ${\cal O}(u^1)$ (giving the linear in $u$, {\bf
$<L>$-Approximation}) and ${\cal O}(u^2)$ (quadratic in $u$,{\bf the
$<Q>$-Approximation}) terms.

\section{$<L>$-A\lowercase{pproximation}}

The first of these, the $<L>$-approximate form for the parametrix is a
product of the $<0>$-approximation parametrix and a linkage operation
factor

\of
\left.
e^{ - i \int_0^s {\delta'_u}^2}
e^{   i  \int_0^t u \cdot a }
\right|_{u \rightarrow 0 }
= e^{ i \int_0^s a^2 -  \frac{i}{s} \left(\int_0^s a\right)^2 }
\zf

\noindent where $a_\mu(t) :\equiv g s \int_{\frac{t}{s}}^1 d\alpha
W_\mu\left(\sigma(\alpha)\right)$ and $W_\mu :\equiv \partial_\mu U$.
Using formulas from Appendix \ref{appendixjedan}, one can obtain: 

\of
\int_0^s a^2 = 2s^3 g^2 \int_0^1 d\alpha \int_0^\alpha d\beta 
\beta W_\mu(\sigma(\alpha))
W_\mu(\sigma(\beta)) 
\zf

\of
\frac1{s} \left(\int_0^s a\right)^2 = 
s^3  g^2 \int_0^1 d\alpha \alpha \int_0^1 d\beta \beta 
W_\mu(\sigma(\alpha)) W^\mu(\sigma(\beta))   
\zf

\noindent so that the final expression for the $<L>$-approximation is:

\of
g^{<L>}(s|x,y|gU) = g^{<0>}(s|x,y|gU) 
e^{2is^3 g^2 \int_0^1 d\alpha  (1-\alpha) 
W_\mu(\sigma(\alpha)) \int_0^\alpha d\beta \beta
W^\mu(\sigma(\beta)) 
} 
\zf

\noindent At $z=0$ it simplifies to:

\of
g^{<L>}(s|x,x|gU) 
= \frac{i}{(4\pi i s)^{\frac{D}2}}
e^{ i s g  U\left(x\right)
 + i \frac{s^3 g^2}{12} W^2(x)
}
\zf

\section{$<Q>$-A\lowercase{pproximation}}

In the case of the $<Q>$-Approximation, one retain up to quadratic terms
(in the $u$ variable) in the Taylor expansion of $U$. The parametrix is
again a product of the $<0>$ form and a factor with the functional
operations:

\of
\left.
e^{ - i \int_0^s {\delta'_u}^2}
e^{   - i  \int_0^s  u \cdot a  +   \frac{i}2 \int_0^s \int_0^s  u 
\cdot {\bf b }\cdot u}
\right|_{u \rightarrow 0 }
= 
e^{-\frac12 \tr\ln({\bf 1}  -  2{\bf \Pi} \cdot {\bf b})
+  i  \int\int a \cdot ({\bf 1}  -  2{\bf \Pi} \cdot {\bf b})^{-1}\cdot
{\bf \Pi} \cdot a}
\zf

\noindent where $b_{\mu\nu}(t,t') :\equiv g s
\int_{\frac{\max{t,t'}}{s}}^1 d\alpha {\cal
H}_{\mu\nu}\left(\sigma(\alpha)\right)$, $\v{{\cal H}}(x):\equiv
(\partial\otimes\partial U)(x)$, and eq. \jna{drugazgodnao} has been used.

The determinantal term $N[gU] :\equiv - \frac12 \tr\ln({\bf 1} - 2{\bf
\Pi} \cdot {\bf b}) = \sum_{n=1}^\infty \frac{2^{n-1}}{n} \tr{\left[ ({\bf
\Pi}\cdot {\bf b})^n\right]}$ can be evaluated term by term, but no
general closed form can be obtained. The first two terms in that sum are
(Appendix \ref{pooosdfishadssssssssssaaaaaa}):

\ofa 
N[gU]_1 &=& + g s^2 \int_0^1 d\alpha \alpha (1-\alpha)
(\partial^2 U)(\sigma(\alpha)) \cr N[gU]_2 &=& 2g^2s^4 \int_0^1
d\alpha (1-\alpha)^2 {\cal H}_{\mu\nu}(\sigma(\alpha)) \int_0^\alpha
d\beta \beta^2
{\cal H}^{\mu\nu}(\sigma(\beta))  
\zffa{n2gu1}

The value of $N[gU]$ at $z=0$
 can be obtained explicitly, using  a different, more direct approach	
(Appendix \ref{pomodabnshjdfvjlkkkkkkkkkk}), with the result: 

\ofa
\left.N[gU]\right|_{z=0}  
&=&
 -\frac12 \tr \ln\left(
\frac{\sin\left(s\sqrt{2g\v{{\cal H}}}
\right)}{s\sqrt{2g{\v{{\cal H}}}}}
\right) = 
 \frac14 \sum_{m=1}^\infty \frac{|B_{2m}|}{m (2m)!} 
(8gs^2)^m {\cal L}_m(x) =
\cr
&=&
 \frac{gs^2}6 {\cal L}_1(x)
+ \frac{g^2s^4}{90} {\cal L}_2(x)
+ \frac{4g^3s^6}{2835} {\cal L}_3(x)
+ \frac{g^4s^8}{4725} {\cal L}_4(x)
+ \frac{16g^5s^{10}}{467 775} {\cal
L}_5(x)
+ \cdots
\zfa

\noindent where ${\cal L}_m(x) :\equiv \tr_L(\v{{\cal H}}^m)$ and $\tr_L$
is the trace over Lorentz indices\footnote{example: $\tr_L {\bf R} :\equiv
R^\alpha{}_\alpha$}.

In a similar way, for $z\ne 0$, one can obtain the $n$th-order
 terms in Taylor expansion of the expression $\chi :\equiv i \int\int a
\cdot ({\bf 1}-2{\bf \Pi} \cdot {\bf b})^{-1}\cdot {\bf \Pi} \cdot a$. The
first two are (Appendix \ref{pekaidedasretneigoraososo}):

{\small

\ofa
\chi_0 &:\equiv & i \int\int a \cdot   {\bf \Pi} \cdot a
= 
i 2s^3 g^2 \int_0^1 d\alpha (1-\alpha) W_\mu(\sigma(\alpha)) 
\int_0^\alpha d\beta \beta W^\mu(\sigma(\beta)) 
\cr
\chi_1 
&:\equiv & 
2 i \int\int a \cdot {\bf \Pi }\cdot {\bf b }\cdot   {\bf \Pi }\cdot a
\cr
&=&
i 2 s^5g^3 \int_0^1 d\alpha \int_0^1 d\beta \int_0^1 d\gamma
(\min(\alpha,\gamma)-\alpha\gamma)
(\min(\beta,\gamma)-\beta\gamma)
W_\mu(\sigma(\alpha)) {\cal H}^{\mu\nu}(\sigma(\gamma))W_\nu(\sigma(\beta))
\zfa 

}

\noindent Notice that retaining only $\chi_0$ is giving the
$<L>$-approximation. 

It is possible to obtain an closed exact form for $\chi$ at $z=0$
(Appendix \ref{itojeproslonazalost}): 

\of 
\left.\chi\right|_{z=0} = \frac{i s g }2
W\cdot
\frac1{{\cal H}} \cdot
\left(
\frac{\tan\left(s\sqrt{\frac{g {\cal H}}2}\right)}
{\left(s\sqrt{\frac{g {\cal
H}}2}\right)}-1\right)\cdot W
\zf

For example, several first members are:

\ofa
\left.\chi_0\right|_{z=0} &=&  i \frac{s^3 g^2}{12} W^2(x)
\cr
\left.\chi_1\right|_{z=0} &=&  i \frac{s^5 g^3}{60} (W\cdot{\cal H}\cdot W)(x)
\cr
\left.\chi_2\right|_{z=0} &=&  i \frac{17}{5040}   g^4 s^7 
(W\cdot{\cal H}^2\cdot W)
\zffa{sofsdhgfghjfkjsgsfdgkj}

The closed exact form of $<Q>$-parametrix is then:

{\small

\ofa
g^{<Q>}(s|x,x|gU) &=& 
i 
\det\left(
\frac{\sqrt{2g{\cal H}}}{4\pi i \sin\left(s\sqrt{2g{\cal H}}\right)}
\right)^{\frac12}
exp\left\{
i \frac12 s g 
W\cdot
\frac1{{\cal H}} \cdot
\left(
\frac{\tan\left(s\sqrt{\frac{g {\cal H}}2}\right)}
{\left(s\sqrt{\frac{g {\cal
H}}2}\right)}-1\right)\cdot W
\right\}
\cr
&=&
i
\frac1{(4\pi i s)^{\frac{D}2}}
e^{ i s g  U + \frac{g s^2}6 {\cal L}_1 + i \frac{s^3 g^2}{12} W^2 +
\frac{g^2s^4}{90} {\cal L}_2 + i \frac{s^5 g^3}{60} (W\cdot{\cal H}\cdot
W) + \frac{4g^3s^6}{2835} {\cal L}_3 + i \frac{17}{5040}   
g^4 s^7 (W\cdot{\cal
H}^2\cdot W) + \frac{g^4s^8}{4725} {\cal L}_4(x) + \cdots}
\zfa

}

\section{T\lowercase{he}
S\lowercase{chwinger}-D\lowercase{e}W\lowercase{itt}
A\lowercase{symptotic} E\lowercase{xpansion of the}
P\lowercase{arametrix}}

\label{SchwingerDeWittAsymptoticExpansionSection}

The Schwinger-DeWitt Asymptotic Expansion \cite{DeWitt} method
 allows an arbitrarily precise approximation of the parametrix
$g(x,y|gU)$. The method is known under many different names (heat-kernel
method, high-temperature asymptotic expansion, etc.) and was applyed in
various problems (spectral geometry of manifolds with boundaries
\cite{Hurt83}, propagation of fields in general relativity \cite{Bunch
78}-\cite{Christensen 78}, etc.).  We are presenting here its simplest
variant with the slight improvement allowed buy one-component nature of
propagating field.

Starting from the Initial Value Problem (IVP) for the parametrix
$g(x,y|gU)$: 

\ofa
-  i\partial_s g(s|x,y|gU) & = & [-\partial_x^2 + gU(x)] g(s|x,y|gU) \cr
g(s=0|x,y|gU) & = & \delta(x-y)
\zfa

\noindent one may solve it in two stages: Firstly for the free case
$U\equiv 0$, with the solution $g_0(s|x,y)  :\equiv
g(s|x,y|0)=\frac{i}{(4\pi s)^{\frac{D}2}} e^{ -i\frac{z^2}{4s}}$.
Secondly, for nonzero potential $U$, introduce the ansatz $g=g_0 \cdot h$.
The IVP for {\bf the reduced Parametrix} $h$ is then:

\ofa
-  i \left(\partial_s + \frac1s z\cdot\partial_x \right) h(s|x,y|gU) 
&=& [-\partial_x^2 + gU(x)]
h(s|x,y|gU) \cr
h(s=0|x,y|gU) & = & 1
\zffa{dadadada}

Although it is not possible to solve the first of these equations exactly,
the method allows one to reach (in principle) any desired precision.  Its
idea is based on the notion that the differential operators on the LHS and
RHS of IVP \jna{dadadada} are homogeneous of degrees $-1$ and $0$
respectively.  Then, for the formal Taylor series $ h(s)=1+
\sum_{n=1}^\infty (is)^n h_n $ (this is an asymptotic series), one will
find a well-separated system (hierarchy) of recursive equations for the
functions $h_n$: 

\ofa
(z\cdot\partial_x + 1) h_1 &=&      gU                      
\cr
(z\cdot\partial_x + 2) h_2 &=&      -\partial_x^2 h_1                       
\cr
(z\cdot\partial_x + 3) h_3 &=&      -\partial_x^2 h_2 + gU h_2
\cr
(z\cdot\partial_x + 4) h_4 &=&      -\partial_x^2 h_3 + gU h_3
\cr
(z\cdot\partial_x + 5) h_5 &=&      -\partial_x^2 h_4 + gU h_4 
\cr 
& \cdots &
\zfa

These can be solved in sequence, beginning with $h_1$, and continued to
any finite order. Regularities will appear, suggesting that a better (in
terms of computational costs) starting point is to use the exponential
form $h=\exp\{\sum_{n=1}^\infty (is)^n k_n\} $. One can always switch from
one form to another, using algebraic relations which connect the two
hierarchies of the coefficient functions $\{h_n(x,y)\}$ and
$\{k_n(x,y)\}$\footnote{ $h_1=k_1$, $h_2=k_2+\frac12 k_1^2$,
$h_3=k_3+k_1k_2 + \frac16 k_1^3$, $h_4=k_4 + k_1k_3 + \frac12 k_2^2
+\frac12 k_1^2 k_2 +\frac1{24}k_1^4$,$\cdots$}.  The hierarchy of
equations for the coefficient functions $k_n$ is:

\ofa
(z\cdot\partial_x + 1) k_1 &=& 	gU			\cr
(z\cdot\partial_x + 2) k_2 &=& 	-\partial_x^2 k_1
\cr
(z\cdot\partial_x + 3) k_3 &=& 	-\partial_x^2 k_2 - (\partial_x k_1)^2
\cr
(z\cdot\partial_x + 4) k_4 &=& 	-\partial_x^2 k_3- 2 \partial_x k_1
\partial_x k_2
\cr
(z\cdot\partial_x + 5) k_5 &=&  -\partial_x^2 k_4 - (\partial_x k_2)^2- 2
\partial_x k_1 \partial_x k_3
\cr
& \cdots &
\zffa{makanepijepivovise} 

One can resolve this hierarchy iteratively up to any desired level of
accuracy ${\cal O}(s^n)$. The first several terms are (Appendix
\ref{sdwizvodjenjaprava}):

\ofa
k_1(x,y) &=& g \int_0^1 d\alpha  U(\sigma(\alpha))
\cr 
k_2(x,y) &=& - g \int_0^1 d\alpha \alpha (1-\alpha) (\partial^2
U)(\sigma(\alpha))
\cr 
k_3(x,y) &=& g \frac12 \int_0^1 d\alpha \alpha^2 (1-\alpha)^2 
(\partial^4 U)(\sigma(\alpha))
+ 
\cr 
& &  \  \ \ \ 
-2 g^2 \int_0^1 d\alpha  (1-\alpha) 
 W(\sigma(\alpha))\int_0^\alpha d\beta \beta 
W(\sigma(\beta) )  
\cr 
k_4(x,y)
&=&
 - \frac16 g \int_0^1 d\alpha \alpha^3 
(1-\alpha)^2 (\partial^6
U)(\sigma(\alpha)) + 
\cr
& & 
+ g^2 \int_0^1 d\alpha \alpha (1-\alpha)^2 (\partial^3
U)(\sigma(\alpha)) \int_0^\alpha d\beta \beta W(\sigma(\beta))
+
\cr
& &
+ 2g^2 \int_0^1 d\alpha (1-\alpha)^2 
{\cal H}_{\mu\nu}(\sigma(\alpha)) \int_0^\alpha d\beta \beta^2
{\cal H}^{\mu\nu}(\sigma(\beta)) 
+
\cr
& &
+ g^2 \int_0^1 \frac{d\alpha}{\alpha} (1-\alpha)^2 
W(\sigma(\alpha)) \int_0^\alpha d\beta \beta^3
(\partial^3
U)(\sigma(\beta)) 
+
\cr
& &
+ g^2 \int_0^1 \frac{d\alpha}{\alpha} (1-\alpha) 
W(\sigma(\alpha)) \int_0^\alpha d\beta \beta^2 [2\alpha
-\beta(1+\alpha)]
(\partial^3
U)(\sigma(\beta)) 
+
\cr
& &
+ g^2 \int_0^1 d\alpha \alpha W(\sigma(\alpha)) 
\int_\alpha^1 d\beta \beta (1-\beta)^2 (\partial^3
U)(\sigma(\beta)) + \cdots
\zfa
 
Although straight-forward, the calculations become very tedious in 
higher orders.

The (exponential-)asymptotic expansion of the parametrix has the approximate
form:

\ofa
g(s|x,y|gU) &=&
\frac{i}{(4\pi is)^{\frac{D}2}} e^{ - i\frac{z^2}{4s}} 
e^{ 
is g\int_0^1 d\alpha U(\sigma(\alpha)) 
+ s^2 g \int_0^1 d\alpha \alpha(1-\alpha)(\partial^2 U)(\sigma(\alpha))} 
\cr 
& & 
\ \ 
e^{ - is^3\left[ \frac12 g \int_0^1
d\alpha \alpha^2 (1-\alpha)^2 (\partial^4 U)(\sigma(\alpha)) -2 g^2
\int_0^1 d\alpha (1-\alpha) W(\sigma(\alpha))\int_0^\alpha
d\beta \beta W(\sigma(\beta)) \right] + s^4k_4 + \cdots}
\zffa{svegatoganebibilodajeperasmestaotisaoupoliciju}

\noindent and one should note the agreement of the $U$-dependent
exponential in that expansion
\jna{svegatoganebibilodajeperasmestaotisaoupoliciju} with terms appearing
in previous approximate forms: for example,
 the $sg$ term comes from the $<0>$-approximation; the $s^3 g^2$ term
comes from the $<L>$-approximation; the $s^2g$ and $s^4g^2$ terms come
from the $<Q>$-approximation ($N_2[gU]_1$ and $N_2[gU]_2$), as will all
higher terms that depend on second derivatives matrix ${\cal H} :\equiv
(\partial \otimes \partial U)$ only (they will be of order $(s^2g)^n \sim
s^{2n}$);  the $<Q>$ correction terms for $<L>$, of the type $W\cdot{\cal
H}^n \cdot W$ ($n>0$) are not visible here, since their order is $s^3g^2
(s^2g)^n \sim s^{3+2n}$.  The term $s^3g$ is proportional to the fourth
derivative (hyper)matrix $(\partial^{\otimes 4}U)$, and cannot come from
either $<L>$ or $<Q>$ approximations. Same is with $ (\partial^{\otimes
6}U)$ ($\sim s^4g$) and higher derivative terms.

The diagonal element (in coordinate space) of the parametrix (i.e. $z=0$,
$\sigma(\alpha)=x$) takes form:

\ofa
g(s|x,x|gU)
&=&
\frac{i}{(4\pi i s)^{\frac{D}2}}
e^{is gU(x) + \frac{s^2 g}6 (\partial^2 U)(x)
- i\frac{s^3 g}{60} \partial^4 U(x) 
+ i \frac{s^3 g^2}{12} W(x)^2
}
\cr
& & 
e^{
-\frac{s^4g}{840} 
(\partial^6U)(x)
+\frac{13 s^4 g^2}{240} 
(\partial^3U)(x)
W(x)
+\frac{s^4g^2}{90}
{\cal H}_{\mu\nu}(x)
{\cal H}^{\mu\nu}(x)
+ \cdots}
\zffa{svegatoganebibilodajeperasmestaotisaoupolicijudijagonala}

\noindent This form can be useful for the evaluation of the functional
determinant that represents vacuum loops in quantum field theory.

As a last point in this section, note that is possible to obtain the full
expression for the collection $F_1(z|s)$ of the first order in $g$ terms
in the exponential $k=\ln h$ (i.e. $h=e^{gF_1(z|s)  +{\cal O}(g^2)}$). 
The result is (Appendix \ref{sdwizvodjenjaperutrjndahsej})

\of
F_1(z|s) = is \int_0^1 d\alpha e^{-is\alpha (1-\alpha) \partial_y^2 }
U(y+\alpha z) = 
is \int_0^1 d\alpha \sum_{n=0}^\infty \left[ 
-is\alpha (1-\alpha) \partial_y^2
\right]^n
U(y+\alpha z)
\zf

\noindent It's diagonal ($z=0$) value is 

\of
F_1(0|s) = is\sum_{n=0}^\infty \frac1{(2n+1)!!}\left(-\frac{is}2
\partial_y^2\right)^n U(y)
= isU + \frac{s^2}6 \partial^2U -\frac{is^3}{60}\partial^4U -\frac{s^4}{840}
\partial^6U + \cdots 
\zf

\noindent which  is in  full agreement with previous derivations.

\section{C\lowercase{onclusion}}

In this paper we have presented a new form of Fradkin's parametrix, as
well as a set of reasonable approximations.

The first of them, the $<0>$-approximation, is eikonal in spirit, in the
sense that it takes into account only the extreme high-energy behaviour of
the propagating particle. It is exact in the case of a laser-like external
field.  That property will be shown (in the next paper)  to be true in
more complex cases (spinor particle, gauge and gravitational external
fields). Since the $<0>$-form has a simple coordinate representation, we
did employ it in calculation of full (nonperturbative) quantum correlation
functions for $\lambda\varphi^4|_{D=4}$ theory.  The effective interaction
vanishes, displaying the "triviality" of theory. 

Beyond the $<0>$-approximation, two directions were explored: first,
$<L>$- and $<Q>$- approximation that are first two corrections to $<0>$,
derived in the spirit of a semiclassical approach; a difference from the
semiclassical approach is that fluctuations are not taken around the
classical trajectories of the particle, but around eikonal ones
(straight-linear trajectories). Another approach is the Schwinger-DeWitt's
asymptotic expansion method (heat kernel method), which we have adapted
and made explicit. The validity of its results is established by a
comparison with $<0|L|Q>$ results.  The majority of longer derivations
have been placed into Appendices.

The author is indebted to H.M.Fried for generous help during both the
research and in preparation of this article.  Thanks are also due to
G.Guralnik, K.Kang, A.Jevicki and P.   
Ne$\v{s}$kovi\'c for useful conversations.
 
This work was suported in part by the U.S. Department of Energy under Contract No.
DE-FG02-91ER40688-Task A.

{
\appendix

\section{S\lowercase{ome useful linkage and functional identities} }

\label{linkageidentiteti}

\of
\left.
e^{ -i\alpha \partial_V^2} f(V) \right|_{V \rightarrow 0} =
\int dV \frac{i}{(4\pi i \alpha)^{\frac{D}2}} 
e^{ - \frac{iV^2}{4\alpha}} f(V)
\zff{prvazgodna}

\of
\left. 
e^{-is\partial_V^2}\delta_{D}(V-a) f(V)\right|_{V\rightarrow 0}=
\frac{i}{(4\pi s)^{\frac{D}2}}
e^{ - \frac{i}{4s}a^2}
f(a) 
\zff{trecazgodna}

\of
\left.  
e^{ - i\int_0^s dt\delta_v^2} F(v) \right|_{v\rightarrow 0} = 
\int
\left[{\cal{D}}v\right] N(s)  e^{ -\frac{i}4 \int_0^s dt v(t)^2} F[v] 
\zff{osmazgodna}

\noindent where $N(s) :\equiv \int \D\eta e^{i\int_0^s \eta^2} = 
\left( {\cal D}v e^{ -\frac{i}4
\int_0^s dt v(t)^2} \right)^{-1}$.

\of
\left.
e^{\int \delta_u \cdot A \cdot \delta_u }
e^{\int \alpha \cdot u + \int u \cdot B \cdot u} 
\right|_{u \rightarrow 0}
= 
\det({\bf 1} - 4AB)^{-\frac12} e^{\int \alpha 
\cdot \frac1{{\bf 1} - 4AB}A \cdot 
\alpha } 
\zff{drugazgodnao}

\section{F\lowercase{unctional form of} G\lowercase{reen's function}}

\label{izvodenjefunkcinalneforme}

 It is possible (using the identity \jna{osmazgodna} ) to write 
\jna{minkowskifradkinpropagator} as a 
functional integral over the auxiliary variable $v(t)$: 

\of
  G(x,y|gU)_c   
= 
i \int_0^\infty ds e^{-ism^2_-} 
\int \left[{\cal D}v\right] 
N(s) 
e^{-\frac{i}4 \int_0^{s} dt v(t)^2} 
e^{i\int_0^{s}   dt g U\left(y + \int_0^{t} v\right)}
 \delta\left(x-y - \int_0^{s} v\right) 
\zff{pathpekatrst}

\noindent where $ N(s)$ is a  normalization constant  fixed by the relation
$ N(s)^{-1}=\int 
\left[{\cal D}v\right]e^{ - \frac{i}4 \int_0^s dt v(t)^2}$.

One can transform the functional-integral form \jna{pathpekatrst}
further into a more-heuristically transparent form,  

\of
  G(x,y|gU)_c = i \int_0^\infty ds e^{-ism^2_-} 
\int DX\D P e^{-i\int_0^{s} dt (P\dot{X} -P^2 -g U(X))}
\delta(X(0)-y)\delta(X(s)-x)
\zff{pathintegralnaforma}

\noindent by introducing an auxiliary variable $X(t)$.
Impose on it 
the constraint $X(t) = y + \int_0^{t-\epsilon} v$, which is
equivalent to the IVP 
$\dot{X}(t)= v(t)$,$X(0)=y$. The idea is to solve that IVP for $v$
(i.e. 
$v(t)= \dot{X}(t)$). Then one can perform a change of variables: 
$v\rightarrow X$ 
by including the identity (see Appendix \ref{pingvinidudidadu})

\of
1 \equiv \int DX \delta\left[X(t) - y  - \int_0^{t-\epsilon} v \right] = 
\int DX\delta\left[\dot{X}(t) - v(t)\right] \delta(X(0)-y)
\zf

\noindent in \jna{pathpekatrst},

\of
  G(x,y|gU)_c
=
i \int_0^\infty ds e^{-ism^2_-}
\int \left[DX \right]
N(s)
\delta(X(0)-y)
\delta(X(s-\epsilon)-x)
e^{i\int_0^{s-\epsilon}   dt g U\left(X(t)\right)}
\
S
\zf

\noindent where the expression $S$ is: 

\ofa
S &\equiv &
\int [Dv]
e^{- \frac{i}4 \int_0^{s} dt v(t)^2}
\delta\left[\dot{X}(t) - v(t)\right]
=
\int [\D P Dv ] e^{-\frac{i}4 \int_0^{s} dt v(t)^2}
e^{-i\int_0^{s} dt [P\dot{X}-Pv]}
=
\int [\D P]
e^{-i\int_0^{s} dt P\dot{X} + i\int_0^{s} dt P^2}
\zfa

Finally, one may assemble all terms to get
\jna{pathintegralnaforma}.

Its heuristic interpretation is very simple; that of a path-integral
amplitude for a particle with Hamiltonian $H(X, P)=P^2+U(X)$, which 
propagates between initial position $y$ and final position $x$. Since
the 
particle is relativistic, the reparametrisational symmetry of its
world line must be fixed, i.e. one should integrate over all possible
choices of world-line length $s$ (between these two points), with the
"weight function" $e^{-ism^2}$ determined by the mass of the 
particle.

\section{C\lowercase{hange of variables}}

\label{pingvinidudidadu}

The proof of the identity:
  
\of
\int DX \delta\left[X(t) - y  - \int_0^{t-\epsilon} v \right] =
\int DX\delta\left[\dot{X}(t) - v(t)\right] \delta(X(0)-y)
\zff{lhsovodoksjsghvuvhb}
 
\noindent may be given as follows.  Start with {\bf the left hand side}
(LHS) of \jna{lhsovodoksjsghvuvhb}, discretize it to

\ofa
LHS &:\equiv& \lim_{N\rightarrow \infty}
\int dX_0 dX_1 \cdot dX_N \delta(X_0-y) \delta(X_1-y -\Delta t v_0)
\cr
& & 
\delta(X_2-y -\Delta t [v_0+v_1]) 
\cdots 
\delta(X_N-y -\Delta t [v_0+ \cdots + v_{N-1}])
\zfa

\noindent where $\Delta t =\epsilon = :\equiv \frac{s}N$, and $X_m:\equiv
X\left(m\Delta t\right)$ and $v_m:\equiv v\left(m\Delta t\right)$.  Note
that the very first $\delta$-function does not have a $v$-variable in its
argument. Then replace $y\rightarrow X_0$ in the second $\delta$-function
(using first $\delta$-function), replace $v_0 \rightarrow
\frac{X_1-X_0}{\delta t}$ in the third $\delta$-function (now using the
second $\delta$-function), etc ... As a result, one obtains

\ofa 
\lim_{N\rightarrow \infty}
(\Delta t)^{N-1}
\int dX_0 dX_1 \cdot dX_N \delta(X_0-y)
\delta\left(\frac{X_1-X_0}{\delta   t} - v_0 \right) 
\cr
\delta\left(\frac{X_2-X_1}{\delta   t} - v_1 \right) 
\cdots 
\delta\left(\frac{X_N-X_{N-1}}{\delta   t} - v_{N-1} \right) 
\zfa

\noindent and then remove the regularization ($N\rightarrow \infty$). The
final result is

\of
\int DX\delta\left[\dot{X}(t) - v(t)\right] \delta(X(0)-y)
\zf
 
\noindent where the $\lim_{N\rightarrow \infty}
(\Delta t)^{N-1} $ is included in the new measure $DX$.

\section{$<0>$-\lowercase{form versus} $<P\lowercase{h}>$-\lowercase{form}}

\label{appendixzanultu}

To transform the $<0>$-form \jna{NULAforma} to \jna{kolikotijegodina},
introduce the auxiliary variable $P:\equiv \frac1s z$ so that

\ofa
g^{<0>}(s|x,y|gU) 
&=& 
\frac{i}{(4\pi i s)^{\frac{D}2}}
e^{ - \frac{iz^2}{4s} + i g \int_0^{s} dt U\left(
\sigma\left(\frac{t}{s}\right)\right) } =
\cr  
&=& 
i(- i \pi  s)^{\frac{D}2}
\int \d P \delta(z-sP) 
e^{ - \frac{isP^2}4  + i g \int_0^{s} dt U(y+tP) } =     
\cr 
&=&
i(- i \pi  s)^{\frac{D}2}
\int \d P \d p e^{ip(z-sP)- \frac{isP^2}4  + i g \int_0^{s} dt U(y+tP) } .
\zfa

\noindent Then represent $i(- i \pi s)^{\frac{D}2}$ as $\int dq
e^{\frac{i}s q^2}$,

\ofa
g^{<0>}(s|x,y|gU)
&=&
\int \d P \d p dq e^{ ipz - ispP - \frac{isP^2}4  
+ \frac{i}s q^2 +
 i g \int_0^{s} dt U(y+tP) }
=
\cr
&=&
\int \d P \d p dq e^{ ipz + isp^2 - \frac{i}4s(P+2p)^2 + \frac{i}s q^2 +
 i g \int_0^{s} dt U(y+tP) } ,
\zfa

\noindent and shift $q\rightarrow q + \frac12s(P+2p)$ to obtain

\ofa
g^{<0>}(s|x,y|gU)
&=&
\int \d P \d p dq e^{ipz + isp^2 + \frac{i}s q^2 + iq(P+2p) +
 i g \int_0^{s} dt U(y+tP) }=   
\cr 
&=& 
\int \d P \d p dq e^{ipz + isp^2 +iq\left(P+2p + \frac1s q\right) +
 i g \int_0^{s} dt U\left(y+tP\right) }
=
\cr
&=&
\int \d P \d p dq e^{ipz + isp^2 - iq(P+2p) +
 i g \int_0^{s} dt U\left(y+tP+ \frac{t}s q\right) } ,
\zfa

\noindent where the last two steps were shifts $P \rightarrow P  - \frac1s q$
and $q \rightarrow -q$.

That is expression \jna{kolikotijegodina}, suitable for comparation with
$<Ph>$-form \jna{malenadevojcice}.

\section{P\lowercase{roof of projected form}}

\label{friedovdokazzamojidentitet}

The easiest way to prove the form \jna{tacniparametrix} is to notice that
constrained variable $\Pi\cdot u = u-u_0$ ($u_0:\equiv \frac1s\int_0^s u$)
is given by the expansion \jna{obivniragdkgm} as $\Pi\cdot
u(t)=\sum_{n=1}^\infty [Q_n^\mu \cos(n\omega t )  + P_n^\mu \sin(n\omega
t)]$. Then $\int_0^s {\delta'}_u^2 = \frac2{s} \sum_{n=1}^\infty \left\{
{\partial^2 \over \partial Q_n^2} + {\partial^2 \over \partial P_n^2}
\right\} $ and $\int_0^t u = t u_0 + \frac{s}{2\pi}
\sum_{n=1}^\infty\frac1{n} \left\{ Q_n \sin(n\omega t) + P_n [1-
\cos(n\omega t)] \right\}$, giving: 

\ofa
g(s|x,y|gU)
&=&
\frac{i}{(4\pi i s)^{\frac{D}2}}
e^{ - \frac{iz^2}{4s} }
 \prod_{n=1}^\infty
e^{ - \frac{2i}{s}
\left\{
{\partial^2 \over \partial Q_n^2} +  {\partial^2 \over \partial P_n^2 }
\right\}
}
\nn
& &
\left.
e^{
i g \int_0^{s} dt
U\left( \sigma\left(\frac{t}s\right)  + t u_0 + \frac{s}{2\pi} 
\sum_{n=1}^\infty
\frac1{n}
        \left\{ Q_n \sin(n\omega t) + P_n [1- \cos(n\omega t)] \right\}
\right)
}
\right|_{\{ Q_n \},\{ P_n \}, u_0\rightarrow 0 }
\zfa

\noindent Since there is no linkage operation
with
respect to $u_0$, one can directly proceed to the 
$u_0\rightarrow 0$ limit, reproducing the 
\jna{tacnamoja} form.

\section{E\lowercase{valuation of} $N\left[\lowercase{g}U\right]_{1,2}$}

\label{pooosdfishadssssssssssaaaaaa}

Start from \jna{345353535353535345434}:

\of 
(\Pi\cdot b)^{\mu t}{}_{\nu \tau} = gs \int_0^1 d\alpha
{\cal H}^\mu{}_\nu 
(\sigma(\alpha))
\left[\theta\left(\alpha-\max\left(\frac{t}{s},\frac{\tau}{s}\right)\right) - 
\alpha\theta\left(\alpha-\frac\tau{s}\right)\right]
\zf

\noindent where ${\cal H}^\mu{}_\nu :\equiv(\partial^\mu\partial_\nu U)$. Then 

\ofa
N\left[gU\right]_1 &=& \tr(\Pi\cdot b) 
= 
gs \int_0^s dt (\Pi\cdot b)^{\mu 
t}{}_{\mu t} 
\cr
&=& 
gs^2 \int_0^1 d\beta  \int_0^1 d\alpha 
(\partial^\mu\partial_\mu U)(\sigma(\alpha))
\left[\theta(\alpha-\beta)- \alpha \theta(\alpha-\beta)\right]
\cr
&=&
gs^2 \int_0^1 d\beta  \int_0^1 d\alpha 
(\partial^2 U)(\sigma(\alpha)) 
(1-\alpha)\theta(\alpha-\beta)
\cr
&=&
gs^2  \int_0^1 d\alpha (\partial^2 U)(\sigma(\alpha))
\alpha(1-\alpha)	
\zfa

\noindent and 

\ofa
N\left[gU\right]_2&=& \tr(\Pi\cdot b \cdot \Pi\cdot b)    
\cr
&=& 
 g^2s^2 \int_0^s dt \int_0^s d\tau 
\int_0^1 d\alpha \int_0^1 d\beta
{\cal H}_{\mu\nu}(\sigma(\alpha)) 
{\cal H}^{\mu\nu}(\sigma(\beta)) 
\cr
& &
\ \ \ \left[\theta\left( 
\alpha-\max\left(\frac{t}{s},\frac{\tau}{s}\right)\right) - 
\alpha 
\theta\left(\alpha-\frac\tau{s}\right)\right]
\cdot
\cr
& &
\ \ \
\cdot
\left[\theta\left(
\beta-\max\left(\frac{t}{s},\frac{\tau}{s}\right)\right)
-
\beta
\theta\left(\beta-\frac{t}{s}\right)\right]
\cr
&=&
g^2s^4 \int_0^1 d\alpha \int_0^1 d\beta 
\int_0^1 d\gamma \int_0^1 d\delta
{\cal H}_{\mu\nu}(\sigma(\alpha))
{\cal H}^{\mu\nu}(\sigma(\beta))
\cr & & 
\ \ \ \left[\theta(\alpha-\max(\gamma,\delta)) -\alpha \theta(\alpha - 
\delta)\right]
\left[\theta(\beta-\max(\gamma,\delta)) -\beta \theta(\beta - \gamma)\right]
\cr
&=&
g^2s^4 \int_0^1 d\alpha \int_0^1 d\beta
\int_0^1 d\gamma \int_0^1 d\delta
{\cal H}_{\mu\nu}(\sigma(\alpha))
{\cal H}^{\mu\nu}(\sigma(\beta))
\cr & &
\ \ \ [\min(\alpha,\beta)^2 - 2\alpha\beta \min(\alpha,\beta)
+\alpha^2\beta^2]
\cr
&=&
2 g^2s^4 \int_0^1 d\alpha 
(1-\alpha)^2 
{\cal H}_{\mu\nu}(\sigma(\alpha))
\int_0^\alpha d\beta \beta^2 
{\cal H}^{\mu\nu}(\sigma(\beta))
\zfa

\noindent where tools from Appendix \ref{appendixjedan}
were utilized.

\section{E\lowercase{valuation of}
$\left.N[\lowercase{g}U]\right|_{\lowercase{z}=0}$}

\label{pomodabnshjdfvjlkkkkkkkkkk}

In this appendix the "loop-generator"  $N[gU] :\equiv -\frac12
\tr\ln({\bf A})$ for matrix ${\bf A}:\equiv {\bf 1} - 2 {\bf \Pi\cdot b}$
 for the case $z=0$ is evaluated. Matrix elements of ${\bf A}$ are
given by

\ofa
A^{\mu t}{}_{\nu \tau} 
&=&
 \eta^\mu{}_\nu \delta(t-\tau) - 2 ({\bf \Pi\cdot b})^{\mu t}{}_{\nu \tau}
 \stackrel{\jna{llllllrrrrrtttttteeeyyyyy}}{=}
=
 \frac1s\left\{\eta^\mu{}_\nu \delta\left(\frac{t}{s}-
\frac{\tau}{s}\right) - gs^2
{\cal H}^\mu{}_\nu(x) 
K\left(\frac{t}{s},\frac{\tau}{s}\right) \right\}
\zfa

\noindent where $K(\alpha,\beta) :\equiv 1- 2 \max(\alpha,\beta) +
\beta^2$. We will exploit fact that $N[gU]$ can be represented in terms of
dimensionless variables ($\alpha=t/s$, etc ...) in the form $N[gU] =
-\frac12\overline{\tr}\ln({\bf \overline{A}})$, where
$\overline{A}^{\mu\alpha}{}_{\nu\beta} = s A^{\mu, t=s\alpha}{}_{\nu,
\tau=s\beta} \ = \eta^\mu{}_\nu \delta(\alpha-\beta) - gs^2 {\cal
H}^\mu{}_\nu(x)  K(\alpha,\beta)$, where $\overline{\tr}(...)= \int_0^1
d\alpha (...)(\alpha,\alpha)$. 
 
The operator ${\hat K}$ whose kernel is $K(\alpha,\beta)$ is not
completely diagonalisable, since it is not a {\bf normal operator}, i.e.
it does not satisfy condition of normality ${\hat
K}^\dagger\hat{K}=\hat{K}\hat{K}^\dagger$. To see this, evaluate matrix
elements of both sides:

\ofa
(\hat{K}^\dagger\hat{K})(\alpha,\gamma)
&=&
\int_0^1 d\beta K(\beta,\alpha) K(\beta,\gamma) 
=
\frac13  -\alpha^2 -\gamma^2 + 2\gamma \alpha^2 +\frac23 \gamma^3
-\alpha^2\gamma^2
\\
(\hat{K}\hat{K}^\dagger)(\alpha,\gamma)
&=&
\int_0^1 d\beta K(\alpha,\beta) K(\gamma,\beta) 
=
\frac15 - \alpha^2 -\gamma^2 + 2\gamma \alpha^2 + \frac23 \gamma^3 
- \frac16 \alpha^4
-\frac16 \gamma^4
\zfa

\noindent where $\gamma \ge \alpha$ is assumed, and
\jna{posaonisam}-\jna{dobioja} were used.

We will here first resolve the decomposition problem of $\hat{K}$, and
only then we will return to evaluation of spectral problem for operator
${\bf \overline{A}}$.

\subsection{Structure of $\hat{K}$}

\label{wwwwuuuuusssssssssshhhhhhkkkkk}

Since $\hat{K}$ is not normal, it cannot be completely diagonalized. 
 Even worst, its left- and right-hand-side eigenvectors are not the same,
and those sets of eigenvectors are not complete basis sets. 

Here, we will find the right-hand-side-eigenvectors of $\hat{K}$. To
 find them, start with the eigen-problem $\int_0^1 d\beta K(\alpha,\beta)
R(\beta)= \lambda R(\alpha)$ and its two differential consequences in
explicit forms: 
 
\of
\int_0^1 d\beta R(\beta)  -2\alpha \int_0^\alpha d\beta R(\beta)
 - 2 \int_\alpha^1
d\beta \beta  R(\beta)  + \int_0^1 d\beta  \beta^2 R(\beta)  
= \lambda R(\alpha)
\zff{ppppppppppppppssssssssss}
 
\of
-2  \int_0^\alpha d\beta R(\beta)  = \lambda R'(\alpha)
\zff{ddddddddddddddoooooooooooo}
 
\of
-2 R(\alpha)  = \lambda R''(\alpha)
\zf
 
Substituting the general solution $R(\alpha) =A e^{i\nu \alpha} + B
e^{-i\nu \alpha}$ of that last equation (where $\lambda=\frac2{\nu ^2}$) 
into \jna{ddddddddddddddoooooooooooo} one gets $B=A$, so $R(\alpha) = 2A
\cos(\nu \alpha)$. Equation \jna{ppppppppppppppssssssssss} further
constrains the solution to frequencies $\nu $ such that $\frac{\sin(\nu
)}{\nu ^2} = 0 $. This gives possible eigen-frequencies $\nu _n \equiv
n\pi$ ($n\ge 1$), corresponding to the eigenvalues $\lambda_n=\frac2{\pi^2
n^2} $ and normalized (in terms of scalar product $<a|b>:\equiv \int_0^1
d\alpha a(\alpha) b(\alpha)$)  eigenvectors $R_n(\alpha)=\sqrt{2}\cos(n\pi
\alpha)$. Notice that the $R_n$ are not the left-hand-side-eigenvectors of
$\hat{K}$.

Also, although the vectors $R_n$ are mutually orthogonal, they are not a
complete set:  the supposed decomposition of unity gives the projection
operator $\sum_{n=1}^\infty R_n(\alpha)  R_n(\beta) =
\delta(\alpha-\beta)-1 \equiv: \pi(\alpha,\beta)$ with defect one.

A similar procedure for left-hand-side-eigenvectors does not allow
solutions in terms of elementary functions, and will not be presented
here. The only left-hand-side-eigenvector that is easy to obtain is
$L_0(\alpha)=1$, with corresponding eigenvalue $0$. It is not the
right-hand-side-eigenvector of $\hat{K}$. 

In terms of ortho-normal basis $\{L_0=1, R_n=\sqrt{2} \cos(n\pi\alpha) ; n
\ge 1\}$ one can write the resolution $\hat{K} = \hat{K}_D + \hat{K}_J$ on
$\hat{K}_D$, the diagonalisable part, and $\hat{K}_J$, the
nondiagonalisable (Jordan type) part:

\ofa
\hat{K}_D 
&=&
\sum_{n=1}^\infty \lambda_n |R_n><R_n| 
\cr
\hat{K}_J
&=&
-\sqrt{2} \sum_{n=1}^\infty (-)^n \lambda_n |R_n><L_0| \equiv: |KL_0><L_0|
\zfa

\noindent where $\lambda_n =\frac{2}{\pi^2 n^2}$, and $|KL_0> :\equiv
\hat{K}|L_0> = -\sqrt{2} \sum_{n=1}^\infty (-)^n \lambda_n |R_n>$ (note
that $<L_0|KL_0> \equiv 0$).

In the above basis, the matrix form of full operator $\hat{K}$ is 

\of
\v{K} = \pmatrix{
0 & 0 & 0 &  0  & 0 & \cdots \cr
\sqrt{2} \lambda_1  & \lambda_1 & 0 & 0 & 0 & \cdots \cr
 -  \sqrt{2} \lambda_2  & 0 & \lambda_2 & 0 & 0 & \cdots \cr
\sqrt{2} \lambda_3  & 0 & 0 & \lambda_3 & 0 & \cdots \cr
-  \sqrt{2} \lambda_4  & 0 & 0 & 0 & \lambda_4 & \cdots \cr
\cdot & \cdot & \cdot & \cdot & \cdot & \cdots \cr
\cdot & \cdot & \cdot & \cdot & \cdot & \cdots \cr
\cdot & \cdot & \cdot & \cdot & \cdot & \cdots \cr
}
\zf

In coordinate form, kernels of the $\hat{K}_{D,J}$ operators are

\ofa
K_J(\alpha,\beta) 
&=&
\int_0^1 d\gamma K(\alpha,\gamma)
=
\frac13 -\alpha^2
\cr
K_D(\alpha,\beta)       
&=&
\int_0^1 d\gamma K(\alpha,\gamma) \pi(\gamma,\beta)    
=
\frac23  - 2 \max(\alpha,\beta) +\alpha^2 + \beta^2
=
\sum_{n=1}^\infty \frac{4}{\pi^2 n^2}\cos(n\pi\alpha)\cos(n\pi\beta)
\zfa

\noindent Also: $KL_0(\alpha) = \frac13 -\alpha^2$.

\subsection{Spectral structure of ${\bf \overline{A}}$; Evaluation of
$\left.N[gU]_2\right|_{z=0}$}

\label{mmmmmmmmeeeeeeeeeiiiiiirrrrrrcccccccu}

Using the known decomposition of the operator $\hat{K}$, one can obtain
the corresponding representation of the operator ${\bf \overline{A}}$: 

\ofa
 {\bf \overline{A}}  
&=&
\hat{1}\otimes{\v{1}}
-gs^2 \hat{K}\otimes{\v{{\cal H}}}
=
\cr
&=&
\left(
|L_0><L_0| + \sum_{n=1}^\infty |R_n><R_n|
\right) \otimes{\v{1}}
-g s^2 \left(
\sum_{n=1}^\infty \lambda_n |R_n> \left\{ <R_n| - \sqrt{2} (-)^n <L_0|\right\}
\right)\otimes{\v{{\cal H}}}
= 
\cr
&=&
\pmatrix{
{\v{1}} & 0 & 0 & 0 & \cdots \cr 
-gs^2 \sqrt{2} \lambda_1 {\v{{\cal H}}} & {\v{1}} 
- gs^2 \lambda_1 {\v{{\cal H}}} & 0 & 0 & \cdots \cr 
gs^2 \sqrt{2} \lambda_2 {\v{{\cal H}}} & 0 & {\v{1}} 
- gs^2 \lambda_2 {\v{{\cal H}}} & 0 &  \cdots \cr 
-gs^2 \sqrt{2} \lambda_3 {\v{{\cal H}}} & 0 & 0 & \v{1} 
- gs^2 \lambda_3 {\v{{\cal H}}} & \cdots \cr 
\cdot & \cdot & \cdot & \cdot & \cdots \cr
\cdot & \cdot & \cdot & \cdot & \cdots \cr
\cdot & \cdot & \cdot & \cdot & \cdots \cr
}
\zffa{spectralstructureofApppiiochj}

From that representation one can read its eigenvalues:

\of
\chi_{a,n}(x) :\equiv 1 - gs^2 \Lambda_a(x) \lambda_n = 1 - \frac{2gs^2
\Lambda_a(x)}{\pi^2 n^2}  \ ; \ \hskip2cm n\ge 1
\zf
 
\noindent where $\Lambda_a(x)$ ($a=\overline{1,D}$) are the eigenvalues of
the matrix $\v{{\cal H}}(x)$. The $n=0^{th}$ eigenvalue is equal $1$.

In  this way,  $\left.N[gU]\right|_{z=0} $ is given as 

\of
\left.N[gU]\right|_{z=0} = -\frac12 \sum_{a=1}^D
\sum_{n=1}^\infty \ln \left(1 - 
\frac{2gs^2 \Lambda_a(x)}{\pi^2 n^2}
\right)
=  -\frac12 \tr_L \ln\left(
\frac{\sin\left(s\sqrt{2g \v{{\cal H}}}\right)}{s\sqrt{2g\v{{\cal H}}}}
\right)
\zf

To obtain the expansion in  $s$, Taylor expand  the logarithm in 
the first form.   The first term is:

\of 
\left.N[gU]_1\right|_{z=0} 
=
 -\frac12 \sum_{a=1}^D \sum_{n=1}^\infty \left(-\frac{2gs^2
\Lambda_a(x)}{\pi^2 n^2} \right) 
= \frac{gs^2}{\pi^2}
\underbrace{\left(\sum_{a=1}^D \Lambda_a(x)\right)}_{{\cal L}_1(x)}
 \underbrace{\left(\sum_{n=1}^\infty \frac1{n^2}\right)}_{S_2}
= \frac{gs^2}6 (\partial^2 U)(x)
\zf

\noindent where $S_2:\equiv \sum_{n=1}^\infty \frac1{n^2}
 = \frac{\pi^2}6$ and $\v{{\cal 
L}}_1(x):\equiv \sum_{a=1}^D \Lambda_a(x) = \tr_L \v{{\cal H}}(x) 
= (\partial^2 U)(x)$. 

The second term in that expansion gives

\of
\left.N[gU]_2\right|_{z=0} =  -\frac12 \sum_{a=1}^D \sum_{n=1}^\infty 
\left(-\frac12\right) \left(-\frac{2gs^2
\Lambda_s(x)}{\pi^2 n^2} \right)^2 = \frac{g^2 s^4}{\pi^4} {\cal L}_2(x) S_4
= \frac{g^2s^4}{90}
{\cal H}^2(x)
\zf

\noindent where ${\cal L}_2(x):\equiv  \sum_{a=1}^D \Lambda_a(x)^2 = 
\tr_L[\v{{\cal H}}(x)^2] 
= {\cal H}_{\mu\nu}{\cal H}^{\mu\nu}$ and $S_4 :\equiv  \sum_{n=1}^\infty
\frac1{n^4} 
= \frac{\pi^4}{90}$.

Similarly, one can evaluate higher terms in 
$\left.N[gU]\right|_{z=0}$,  to obtain 

\ofa
\left.N[gU]\right|_{z=0} &=&  \frac14 \sum_{m=1}^\infty
\frac{|B_{2m}|}{m (2m)!} 
(8gs^2)^m {\cal L}_m(x)
\cr
&=&
 \frac{gs^2}6 {\cal L}_1(x)
+ \frac{g^2s^4}{90} {\cal L}_2(x)
+ \frac{4g^3s^6}{2835} {\cal L}_3(x)
+ \frac{g^4s^8}{4725} {\cal L}_4(x)
+ \frac{16g^5s^{10}}{467 775} {\cal L}_5(x)
+ \cdots
\nn
\zfa

\noindent where ${\cal L}_m(x) :\equiv \tr_L[{\cal H}(x)^m]$.

\section{O\lowercase{n evaluation of} $\chi[\lowercase{g}U]$}

\label{pekaidedasretneigoraososo}

Starting from

\ofa
a_\mu(t) &=& gs \int_{t/s}^1 d\alpha W_\mu(\sigma(\alpha))
\cr
b_{\mu\nu}(t_1,t_2) &=& gs
\int_{\max\left(\frac{t_1}{s},\frac{t_2}{s}\right)}^1 d\alpha
{\cal H}_{\mu\nu}(\sigma(\alpha))
\zfa

and $ \Pi(t_1,t_2) = \delta(t_1-t_2) -\frac1s $,  we obtain:

{\small

\ofa
(a_{\perp})_\mu(t) &:\equiv & (a\cdot\Pi)_\mu(t) 
=
 gs \int_0^1 d\alpha W_\mu(\sigma(\alpha)) 
\left[
\theta\left(\alpha - \frac{t}{s}\right) - \alpha 
\right]
\cr
(b_{\perp\perp})_{\mu\nu}(t_1,t_2) &:\equiv& (\Pi\cdot
b\cdot\Pi)_{\mu\nu}(t_1,t_2) 
= gs \int_0^1 d\alpha
{\cal H}_{\mu\nu}(\sigma(\alpha))
\left[\theta\left(\alpha - \frac{t_1}{s}\right) - \alpha\right]
\left[\theta\left(\alpha - \frac{t_2}{s}\right) - \alpha\right]
\zfa

}
 
Then 

{\small

\ofa
\chi_0 
&=& 
i \int a_\perp \cdot  a_\perp 
= i \int_0^s dt g^2s^2 
\int_0^1d\alpha\int_0^1d\beta W_\mu(\sigma(\alpha)) W^\mu(\sigma(\beta)) 
\left[\theta\left(\alpha-\frac{t}{s}\right)-\alpha\right]
\left[\theta\left(\beta-\frac{t}{s}\right)-\beta\right]
=
\cr
&=&
i g^2s^3 \int_0^1d\alpha\int_0^1d\beta\int_0^1d\gamma W_\mu(\sigma(\alpha))
W^\mu(\sigma(\beta)) 
\left[\theta\left(\alpha-\gamma\right)-\alpha\right]
\left[\theta\left(\beta-\gamma\right)-\beta\right]
=
\cr
&\stackrel{\jna{ionaisto}}{=}& 
i g^2s^3 \int_0^1d\alpha\int_0^1d\beta W_\mu(\sigma(\alpha))    
W^\mu(\sigma(\beta))
\left[
\min(\alpha,\beta) - \alpha\beta
\right]
=
\cr
&=&
 2 i g^2s^3 \int_0^1d\alpha\int_0^\alpha d\beta W_\mu(\sigma(\alpha))
W^\mu(\sigma(\beta))
\left[
\min(\alpha,\beta) - \alpha\beta
\right]
\zfa

}

On the same way:

{\small

\ofa 
\chi_1 
&=& 
2i \int\int a_\perp \cdot b_{\perp\perp} \cdot a_\perp 
= 
\cr
&=&
2 i \int_0^s dt_1 \int_0^s dt_2 g^3s^3 \int_0^1d\alpha \int_0^1d\beta
\int_0^1d\gamma W_\mu(\sigma(\alpha))  {\cal H}^{\mu\nu}(\sigma(\gamma))
W_\nu(\sigma(\beta)) \cdot
\cr
&&
\ \ \ 
\cdot \left[\theta\left(\alpha-\frac{t_1}{s}\right)-\alpha\right]
\left[\theta\left(\beta-\frac{t_2}{s}\right)-\beta\right]   
 \theta\left(\gamma - \max\left(\frac{t_1}{s},\frac{t_2}{s}\right)\right)
= 
\cr
&=&
2 i g^3 s^5 \int_0^1d\alpha \int_0^1d\beta   
\int_0^1d\gamma W_\mu(\sigma(\alpha))  {\cal H}^{\mu\nu}(\sigma(\gamma))
W_\nu(\sigma(\beta))
\cdot\cr&& \ \ \cdot\int_0^1d\mu \int_0^1d\nu
\left[\theta\left(\alpha-\mu\right)-\alpha\right]
\left[\theta\left(\beta-\nu\right)-\beta\right]
 \theta\left(\gamma -\max(\mu,\nu)\right)
=
\cr
&\stackrel{\jna{ionimisvi}, \jna{ionaisto}}{=}&
2 i g^3 s^5 \int_0^1d\alpha \int_0^1d\beta
\int_0^1d\gamma W_\mu(\sigma(\alpha))  {\cal H}^{\mu\nu}(\sigma(\gamma))
W_\nu(\sigma(\beta))
(\min(\alpha,\gamma) - \alpha\gamma) 
(\min(\beta,\gamma) - \beta\gamma) 
\zfa

}

At $z= 0$ the $\alpha$,$\beta$ and $\gamma$ integrals can be performed
explicitely, using formulas from Appendix \ref{appendixjedan}, to obtain same
results as
in 
Appendix\ref{itojeproslonazalost}. However, derivation of $\chi_m$'s (at $z=0$)
given there is superior to
this one (much less tedious).

\section{E\lowercase{valuation of}
$\left.\chi[\lowercase{g}U]\right|_{\lowercase{z}=0}$}

\label{itojeproslonazalost}

From \jna{spectralstructureofApppiiochj} is easy to obtain the inverse of
operator ${\bf \overline{A}}$ as: 

\ofa
{{\bf \overline{A}}}^{-1} 
&=&
|L_0><L_0|\otimes{\v{1}} 
+ 
\sum_{n=1}^\infty 
|R_n><R_n|
\otimes
(\v{1} - gs^2\lambda_n {\v{{\cal H}}})^{-1}
+
\cr
& &
-
\sum_{n=1}^\infty 
\sqrt{2} (-)^n g s^2 \lambda_n
|R_n><L_0|
\otimes
(\v{1} - gs^2\lambda_n {\v{{\cal H}}})^{-1}
{\v{{\cal H}}}
\zfa

To evaluate $\chi$, only projection of that inverse is necessary:

\ofa
{{\bf \overline{A}}}^{-1} \cdot {{\bf \pi}}
&=&
\sum_{n=1}^\infty
|R_n><R_n|
\otimes
(\v{1} - gs^2\lambda_n {\v{{\cal H}}})^{-1}
\zfa

\noindent where ${{\bf \pi}} = \sum_{n=1}^\infty |R_n><R_n| \otimes \v{1}$.
To obtain ${\bf A}^{-1} \cdot {{\bf \Pi}}$ multiply above result with $s$.
Then

\ofa
\left.\chi\right|_{z=0} 
&=& 
i \int\int a\cdot {{\bf A}}^{-1} \cdot {{\bf \Pi}} \cdot a 
=
is \int_0^1 d\alpha \int_0^1 d\beta 
\sum_{n=1}^\infty R_n(\alpha) R_n(\beta) 
a^\mu(s\alpha)
\left(\frac1{ \v{1} - gs^2\lambda_n {\cal H}(x)}\right)_{\mu\nu}
a^\nu(s\beta) 
\zfa

\noindent where $a^\mu(s\alpha) = gs (1-\alpha)W^\mu(x)$, $W^\mu(x):\equiv
\partial^\mu U(x)$. Then

\ofa
\left.\chi\right|_{z=0}
&=&
2 i s^3 g^2 \sum_{n=1}^\infty \left[\int_0^1 d\alpha
(1-\alpha)\cos(n\pi\alpha)\right]^2 W\cdot(\v{1} - gs^2\lambda_n {\cal 
H}(x))^{-1}\cdot W 
=
\cr
&=&
\frac{2 i s^3 g^2}{\pi^4}  \sum_{n=1}^\infty \frac{[1-(-)^n]^2}{n^4} 
W\cdot(\v{1}
- gs^2\lambda_n {\cal
H}(x))^{-1}\cdot W
\zffa{ohohoalajatovolim}

From this point one can proceed in two alternate routs. First is to Taylor-expand
matrix factor in the above form and to obtain an explicit expressions for
$\chi_m$'s
($\chi=\sum_{m=0}^\infty \chi_m$):

\of
\left.\chi_m\right|_{z=0} = i \frac{2^{m+1}s^{2m+3}g^{m+2}}{\pi^{2m+4}}
(W\cdot{\cal H}^m\cdot W) \v{S}_{2m+4}
\zf

\noindent where 

\of
\v{S}_{2m+4} \equiv  \sum_{n=1}^\infty \frac{[1-(-)^n]^2}{n^{2m+4}} = 4
\left(1-2^{-2m-4}\right)
\zeta(2m+4) = \frac{2\left(2^{2m+4}-1\right)\pi^{2m+4}}{(2m+4)!} |B_{2m+4}|
\zf

\noindent giving

\of
\left.\chi_m\right|_{z=0} = i
\frac{2^{m+2}\left(2^{2m+4}-1\right)s^{2m+3}g^{m+2}|B_{2m+4}|}{(2m+4)!}
(W\cdot{\cal H}^m\cdot W)
\zf

Several first members of that series are:

\ofa
\left.\chi_0\right|_{z=0} 
&=&
\frac{i \  s^3 \ g^2 \ W^2}{12}
\cr
\left.\chi_1\right|_{z=0}
&=&
\frac{i \  s^5 \  g^3 \  W\cdot{\cal H}\cdot W}{60}
\cr 
\left.\chi_2\right|_{z=0}
&=&
\frac{ i \ 17 \ s^7 \ g^4 \ W\cdot{\cal H}^2\cdot W}{5040}
\cr
\left.\chi_3\right|_{z=0}
&=&
\frac{ i \ 31 \ s^9 \  g^5 \ W\cdot{\cal H}^3\cdot W}{37800}
\zfa

The second possible route is to regroup terms in \jna{ohohoalajatovolim}:

\of
\left.\chi\right|_{z=0}
=
\frac{2 i s^3 g^2}{\pi^4}  W\cdot \sum_{n=1}^\infty \frac{[1-(-)^n]^2}{n^2}
\frac1{n^2-a^2}
\cdot W
\zf

\noindent where $a^2:\equiv \frac{2gs^2}{\pi^2}{\cal H}$, and to evaluate
given sum

\of
\sum_{n=1}^\infty \frac{[1-(-)^n]^2}{n^2}
\frac1{n^2-a^2} 
=
\frac{\pi^2}{2a^2} \left( \frac{\tan\left(\frac{\pi
a}{2}\right)}{\left(\frac{\pi a}{2}\right)}-1\right)
\zf
 
Then

\of 
\left.\chi\right|_{z=0}
=
\frac{i s^3 g^2}{\pi^2}  
W\cdot 
\frac1{a^2} 
\left( 
\frac{\tan\left(\frac{\pi a}2 \right)}
{\left(\frac{\pi a}2\right)}-1
\right)
\cdot W
= 
\frac{i s g }2 
W\cdot 
\frac1{{\cal H}} \cdot 
\left(
\frac{\tan\left(s\sqrt{\frac{g {\cal H}}2}\right)}
{\left(s\sqrt{\frac{g {\cal
H}}2}\right)}-1\right)\cdot W
\zf

\section{T\lowercase{ools for parametrix calculations}}
\label{appendixjedan}

{ 
\small

\ofa 
(z.\partial_x+n) k_n(x,y) = j_n(x,y) 
&\Rightarrow&  
k_n(x,y) = \int_0^1 d\alpha \alpha^{n-1} j_n(\sigma(\alpha),y) 
\label{pipikvakva}
\\
\partial_x^m f(\sigma(\alpha)) 
&=& 
\alpha^m (\partial f)(\sigma(\alpha))
\\
\int_0^1 d\alpha \alpha^m f(\sigma(\alpha\beta) ) 
&=& 
\beta^{-m-1}\int_0^\beta d\alpha \alpha^m f(\sigma(\alpha))
\\           
\int_0^1 d\beta \beta^l  \int_0^1 d\alpha \alpha^m f(\sigma(\alpha \beta)  )
&=&
\frac1{l-m} \int_0^1 d\alpha 
(\alpha^m-\alpha^l) f(\sigma(\alpha))
\\
\int_0^1d\beta \beta^n\int_0^1 d\alpha f(\alpha\beta) 
&=&
 \frac1{n} 
\int_0^1 d\alpha (1-\alpha^n) f(\alpha)
\\
\int_0^1 d\gamma \gamma^l \int_0^1 d\alpha \alpha^m
\int_0^1 d\beta \beta^n f(\alpha\gamma) g(\beta\gamma) 
 &=& 
 \frac1{l-m-n-1}
\int_0^1 d\alpha \alpha^m f(\alpha) \left\{ (1-\alpha^{l-m-n-1})
\int_0^\alpha d\beta \beta^{{\bf n}} g(\beta) 
\right.
\cr 
& & 
\ \ \ \ + 
\left.
\int_\alpha^1  d\beta \beta^m (1-\beta^{l-m-n-1}) g(\beta)
\right\}
\\
\int_0^1 d\gamma \gamma^l \int_0^1 d\alpha \alpha^m \int_0^\alpha
d\beta 
\beta^n
f(\alpha\gamma) g(\beta\gamma) &=&
 \frac1{l-m-n-1}  \int_0^1 d\alpha 
(\alpha^m-\alpha^{l-n-1})f(\alpha) 
\int_0^\alpha d\beta \beta^n g(\beta)
\\
\int_0^1 d\alpha f(\alpha) \int_\alpha^1 d\beta g(\beta) 
&=& 
\int_0^1 d\alpha g(\alpha) \int_0^\alpha d\beta f(\beta)
\\
\int_0^1 d\beta \Theta(\alpha-\max(\beta,\gamma)) 
&=& 
\alpha \Theta(\alpha-\gamma)
\\
(\Pi\cdot b)_{\mu\nu}(t_1,t_2) 
&=&
 gs
\left[
\int_{\max(t_1,t_2)/s}^1 d\alpha (\partial_\mu \partial_\nu U)(\sigma(\alpha))
- \int_{t_2/s}^1 d\alpha \alpha  (\partial_\mu \partial_\nu U)(\sigma(\alpha))
\right]
\label{345353535353535345434}
\\
\left.(\Pi\cdot b)_{\mu\nu}(t_1,t_2)\right|_{z=0}
&=&
 gs
\left[
\frac12 - \max\left(\frac{t_1}{s},\frac{t_2}{s}\right) 
+ \frac12\left(\frac{t_2}{s}\right)^2
\right]
{\cal H}_{\mu\nu}(x) = \frac12 gs {\cal H}_{\mu\nu}(x) 
K\left(\frac{t_1}{s},\frac{t_2}{s}\right)
\label{llllllrrrrrtttttteeeyyyyy}
\\   
\int_0^1 d\gamma \int_0^1 d\delta 
\theta(\alpha - \max(\gamma,\delta)) 
\theta(\beta - \max(\gamma,\delta)) 
&=& 
\min(\alpha,\beta)^2
\label{677777666666666666}
\\
\int_0^1 d\gamma \int_0^1 d\delta
\theta(\alpha - \max(\gamma,\delta))
\theta(\beta - \gamma) 
&=&
 \alpha \min(\alpha,\beta)
\label{68888888889999}
\\
\int_0^1 d\beta \beta^n \max(\beta, \alpha) 
&=&
\frac1{(n+2)}\left(1+\frac1{n+1}\alpha^{n+2}\right)
\label{posaonisam}
\\
\int_0^1 d\beta\max(\beta, \alpha)
&=& 
\frac12\left(1+\alpha^2\right)
\\
\int_0^1 d\beta \beta^2 \max(\beta, \alpha)
&=&
\frac14\left(1+\frac13\alpha^4\right)
\\
\int_0^1 d\beta \max(\beta, \alpha) \max(\beta, \gamma)
&=&
 \frac13 + \frac12\gamma\alpha \min(\gamma,\alpha) + \frac16
\max(\gamma,\alpha)^3
\label{dobioja}
\\
\int_0^1d\gamma \left[\theta\left(\alpha-\gamma\right)-\alpha\right]
\left[\theta\left(\beta-\gamma\right)-\beta\right]
 &=&
\min(\alpha,\beta) - \alpha\beta
\label{ionaisto}
\\
\theta\left(\gamma -\max(\mu,\nu)\right) 
&=& 
\theta\left(\gamma -\mu\right) \theta\left(\gamma -\nu\right)
\label{ionimisvi}
\\
\int_0^1d\beta (\min(\alpha,\beta) - \alpha\beta ) 
&=& 
\frac12\alpha(1-\alpha)
\label{alineiaca}
\zfa

}

\section{S\lowercase{chwinger}-D\lowercase{e}W\lowercase{itt calculations}}

\label{sdwizvodjenjaprava}

Rewrite  the system \jna{makanepijepivovise} in the form 
 
\of
(z\cdot\partial_z + n) k_n[y,z] = j_n[y,z]
\zf

\noindent where $n \ge 1$, $k_n[y,z]:\equiv k_n(x,y)$ and:  $j_1 = gU$,
$j_2= -\partial_z^2 k_1$, $j_3 = - \partial_z^2 k_2 - (\partial_z k_1)^2$,
$j_4 = - \partial_z^2 k_3 -2 (\partial_z k_1) (\partial_z k_2)$, $\cdots$. 
Then, for every given (known) $j_n[y,z]$, the corresponding $k_n[y,z]$ is
given by the expression (Appendix \ref{appendixjedan}, Eq\jna{pipikvakva})

\of
k_n[y,z] = \int_0^1
d\alpha \alpha^{n-1} j_n[y,\alpha z]
\zf

Then, with known $k_n[y,z]$, one can evaluate the next source function
$j_{n+1}[y,z]$, etc ... During the calculations several characteristic
procedures are taking place. As an illustration, let us calculate the
first several $k$s. 

Since $j_1[y,z] = gU(y+z)$, it follows that $k_1[y,z] = g \int_0^1 d\alpha
U(y+\alpha z)$. 

The second source function takes the form $j_2[y,z]= - g \partial_z^2
\int_0^1 d\alpha U(y+\alpha z)$. Then, each $\partial_z$ converts into
$\alpha \partial_y$ upon action on $U(y+\alpha z)$.  So $j_2[y,z]= -g
\int_0^1 d\alpha \alpha^2 (\partial^2 U)(y+\alpha z)$. That was first
characteristic procedure: $\partial_z \rightarrow \alpha \partial_y$.

Then $k_2[y,z] = \int_0^1 d\alpha \alpha j_2[y,\alpha z] = -g \int_0^1
d\alpha \alpha\int_0^1 d\beta \beta^2 (\partial^2 U)(y+\beta \alpha z)$. 
Rescale $\beta \rightarrow \beta/\alpha$ to make argument of $U$ free of
$\alpha$:  $k_2[y,z] = -g \int_0^1 d\alpha \alpha \int_0^\alpha d\beta
\beta^2 \alpha^{-3} (\partial^2 U)(y+\beta z) = -g \int_0^1 d\alpha
\alpha^{-2} \int_0^\alpha d\beta \beta^2 (\partial^2 U)(y+\beta z)$ (The
second procedure: $\beta \rightarrow \beta/\alpha$). Extend the range of
integration of $\beta$ back to $[0,1]$, introducing the step function
$\theta(\alpha-\beta)$, switch the order of integration (first to perform
the $\alpha$ integral, and only then the $\beta$ integral), and limit the
range of integration of $\alpha$ to $[\beta,1]$ (according to the argument
of the $\theta$ function):  $k_2[y,z] = -g \int_0^1 d\beta \beta^2
(\partial^2 U)(y+\beta z) \int_\beta^1 d\alpha \alpha^{-2}$ (Third
procedure: switch the order of integrals). The integral over $\alpha$ can
be performed, giving $\int_\beta^1 d\alpha \alpha^{-2} =
\beta^{-1}(1-\beta)$ and $k_2[y,z] = -g \int_0^1 d\beta
\beta(1-\beta)(\partial^2 U)(y+\beta z)$. 
 
The third source function becomes $j_3[y,z]= - \partial_z^2 k_2 -
(\partial_z k_1)^2 = g \int_0^1 d\beta \beta^3(1-\beta)(\partial^4
U)(y+\beta z) - g^2 \int_0^1 d\beta \beta (\partial U)(y+\beta z) \int_0^1
d\gamma \gamma (\partial U)(y+\gamma z)$.  Replaying the same order of
procedures as above, one can obtain (in seven lines of derivation) 
$k_3[y,z] = \frac{g}2 \int_0^1 d\alpha \alpha^2 (1-\alpha)^2
(\partial^4U)(\sigma(\alpha)) - 2 g^2 \int_0^1 d\alpha (1-\alpha)
(\partial U)(\sigma(\alpha)) \int_0^\alpha d\beta \beta (\partial
U)(\sigma(\beta))$.

In the same way, the fourth $k$ function can be obtained as: 

\ofa
k_4[y,z] 
&=&
 - \frac16 g \int_0^1 d\alpha \alpha^3 (1-\alpha)^2 (\partial^6
U)(\sigma(\alpha)) + 
\cr
& & 
+ g^2 \int_0^1 d\alpha \alpha (1-\alpha)^2 (\partial^3
U)(\sigma(\alpha)) \int_0^\alpha d\beta \beta (\partial U)(\sigma(\beta))
+
\cr
& &
+ 2g^2 \int_0^1 d\alpha (1-\alpha)^2 (\partial_\mu\partial_\nu
U)(\sigma(\alpha)) \int_0^\alpha d\beta \beta^2 (\partial^\mu\partial^\nu 
U)(\sigma(\beta)) 
+
\cr
& &
+ g^2 \int_0^1 \frac{d\alpha}{\alpha} (1-\alpha)^2 (\partial
U)(\sigma(\alpha)) \int_0^\alpha d\beta \beta^3 (\partial^3
U)(\sigma(\beta)) 
+
\cr
& &
+ g^2 \int_0^1 \frac{d\alpha}{\alpha} (1-\alpha) (\partial
U)(\sigma(\alpha)) \int_0^\alpha d\beta \beta^2 [2\alpha -\beta(1+\alpha)]
(\partial^3
U)(\sigma(\beta)) 
+
\cr
& &
+ g^2 \int_0^1 d\alpha \alpha (\partial
U)(\sigma(\alpha)) \int_\alpha^1 d\beta \beta (1-\beta)^2 (\partial^3
U)(\sigma(\beta)) 
\nn
\zfa 

\noindent This takes about $30$ lines (one and one-half pages of
calculation), that is a factor of four longer than the evaluation of
$k_3$. One can expect that the length of computation for higher order
functions will grow at least as a geometric progression.  It is
interesting that all steps (procedures) are straightforward, and one could
be tempted to program some symbolic mathematical tool to perform them.

Some of the time-saving formulas obtained during calculations are placed
in Appendix \ref{appendixjedan}. 

It is interesting that if one asks different kind of questions, then
certain exact (i.e. in all orders in $s$) partial information can be
obtained. For example, in the Appendix \ref{sdwizvodjenjaperutrjndahsej}
the first order in $g$ contribution to $k$ is derived. 

\section{D\lowercase{erivation of} $F_1(\lowercase{z|s})$}
 
\label{sdwizvodjenjaperutrjndahsej}

From  \jna{dadadada} follows the IVP for $F_1(z|s)$:

\ofa
\left[
-i(s\partial_s + z\cdot\partial_z ) + s\partial_z^2 
\right] F_1(z|s) &=& s U(x)
\cr
F_1(z|0) &=& 0
\zffa{pomocnikcarobnjakovisehfgfgfgf} 

Then, using Schwinger's parametrization one can write

\of
F_1(z|s) = i \int_0^\infty dt e^{- i t \left[
-i( s\partial_s +  z\cdot\partial_z ) + s \partial_z^2 
\right]} s  U(x)
\zf 

For a moment switch to an algebraic structure by introducing differential
generators $\hat{a}:\equiv -s \partial_s$, $\hat{b}:\equiv -z\cdot
\partial_z$ and $\hat{c}:\equiv -is\partial_z^2$ with the commutational
relations $[\hat{a},\hat{b}]=0$, $[\hat{a},\hat{c}]= - \hat{c}$,
$[\hat{a},\hat{b}]= 2\hat{c}$. Then we are interested in $\hat{r}:\equiv
e^{t(\hat{a} + \hat{b} + \hat{c})}$. Assume that it can be factorized in
form $\hat{r} = e^{A(t) \hat{a}} e^{B(t) \hat{b}}e^{C(t) \hat{c}}$. Then
the \jna{pomocnikcarobnjakovisehfgfgfgf} gives equation

$$ \hat{a} + \hat{b} + \hat{c} = \dot{A} \hat{a} + \dot{B} e^{A
\Ad(\hat{a})}\hat{b} + \dot{C} e^{A \Ad(\hat{a})} e^{B
\Ad(\hat{b})}\hat{c} $$

\noindent Since $e^{A \Ad(\hat{a})}\hat{b} = \hat{b}$, $e^{B
\Ad(\hat{b})}\hat{c} = e^{2B} \hat{c}$ and $e^{A \Ad(\hat{a})}\hat{c} =
e^{-A} \hat{c}$, the IVPs for $A$, $B$ and $C$ follow:
$\dot{A}=\dot{B}=1$, $\dot{C}=e^{A-2B}$, $A(0)=B(0)=C(0)=0$. Their
solution is $A(t)=B(t)=t$, $C(t)=1-e^{-t}$ giving:  $\hat{r} =
e^{t\hat{a}}e^{t\hat{b}}e^{(1-e^{-t})\hat{c}}$. So: 

\of   
F_1(z|s) = i \int_0^\infty dt
 e^{ - t s\partial_s}
 e^{ -t z\cdot\partial_z }
 e^{ -i  s (1-e^{-t})\partial_z^2 }
 s  U(x)
\zf

Use now the Fourier representation for $U(x)$ ($x=y+z$) to obtain:

\ofa
F_1(z|s)
& = &
\int \d p e^{ipy} 
 i \int_0^\infty dt
 e^{ - t s\partial_s}
 e^{ -t z\cdot\partial_z }
e^{ipz}
 e^{ +i  s (1-e^{-t})p^2 }
 s \tilda{U}(p)
\cr
& = &
\int \d p e^{ipy}
i \int_0^\infty dt
 e^{ - t s\partial_s}
e^{ipz e^{-t}}
e^{ +i  s (1-e^{-t})p^2 }        
 s \tilda{U}(p)
\cr
& = &
\int \d p e^{ipy}
i \int_0^\infty dt e^{ipz e^{-t}}
e^{ +i  s e^{-t} (1-e^{-t})p^2 }
e^{-t}  s \tilda{U}(p)
\cr
& = &
i s \int_0^\infty dt e^{-t}
\int \d p 
e^{ip \left(y + z e^{-t}\right) +i  s e^{-t} (1-e^{-t})p^2}
\tilda{U}(p)
\zfa

\noindent i.e. 

\of
F_1(z|s) = i s \int_0^\infty dt e^{-t} \sum_{n=0}^\infty 
\frac1{n!} 
\left(-i  s e^{-t} (1-e^{-t}) \partial_y^2\right)^n U\left(y + z e^{-t}\right)
\zf

Introduce $\alpha=e^{-t}$ and the final expression is established:

\ofa
F_1(z|s) 
&=&
 i s \int_0^1 d\alpha  \sum_{n=0}^\infty
\frac1{n!}
\left(-i  s \alpha(1-\alpha) \partial_y^2\right)^n U(y + z \alpha)
\cr
&=&
is\int_0^1 d\alpha U(\sigma(\alpha))
+ s^2 \int_0^1 d\alpha \alpha(1-\alpha) \partial^2U(\sigma(\alpha))
+ 
\cr
& & 
-\frac{is^3}2 \int_0^1 d\alpha \alpha^2(1-\alpha)^2 
\partial^4U(\sigma(\alpha))
- \frac{s^4}6 \int_0^1 d\alpha \alpha^3(1-\alpha)^3 
\partial^6U(\sigma(\alpha)) + \cdots
\nn
\zfa

It is easy to check that first four terms indeed do agree with expressions
obtained through more tedious calculations in previous Appendix.

At $z=0$ the $\alpha$ dependence disappears from argument of $U$ and
integrals
 over $\alpha$ can be performed ($\int_0^1 d\alpha \alpha^n(1-\alpha)^n = B(n+1,n+1) =
\frac{n!n!}{(2n+1)!}$), giving 

\ofa
F_1(z|0) &=& is  \sum_{n=0}^\infty \frac1{(2n+1)!!} 
\left(\frac{-is}{2}\partial_y^2\right)^n U(y)
\cr
&=& 
is U(y) + \frac{s^2}6 (\partial^2 U)(y)
- \frac{is^3}{60} (\partial^4 U)(y)
- \frac{s^4}{840} (\partial^6 U)(y) + \cdots 
\zfa

}

\end{document}